# Solving the four-dimensional $NN - \pi NN$ equations for scalars below meson-production threshold

D. R. Phillips[*] and I. R. Afnan[†]

*Department of Physics, The Flinders University of South Australia, GPO Box 2100, Adelaide 5001, Australia*

(October 3, 2018)

The four-dimensional $NN - \pi NN$ equations are adapted to the case of scalar particles with a $\phi^2 \sigma$ interaction Lagrangian, and solved for energies below the $\sigma$-production threshold. This is achieved in the approximation where $\phi \sigma$ scattering is dominated by the $s$-channel $\phi$-pole term. The importance of the removal of double-counting is investigated, and a detailed comparison of the results of a covariant coupled-channels formulation and the Bethe-Salpeter equation in the ladder and ladder plus crossed-box approximations is presented. A brief discussion of the extension of the method to energies above the $\sigma$-production threshold is given.

## I. INTRODUCTION

Quantum Chromodynamics (QCD) is now the accepted theory of the strong interaction. Its highly non-perturbative character at hadronic mass scales means that effective field theories of the strong interaction are almost always used when the properties of few-hadron systems are calculated. In some such field theories (e.g. those based on Chiral Perturbation Theory) the Lagrangian is organized so that a perturbative calculation of hadronic reactions is appropriate, at least for "small" momenta [1–3]. However, to examine the scale at which quark-gluon degrees of freedom become essential we need to use these field theories to calculate hadronic reactions at medium to large momenta. The need for explicit quark-gluon degrees of freedom may then be deduced from failures of effective hadronic field theories to explain the experimental data. It is in this regime that Chiral Perturbation Theory fails, and unitarity plays an important role.

The amplitudes for any hadronic reaction are, of course, exactly given by the infinite hierarchy of Schwinger-Dyson equations, which can be derived by functional techniques [4–6]. An alternative method of deriving a set of coupled equations for the amplitudes of a field theory was developed by Taylor [7,8]. It involves taking the series of Feynman diagrams for a particular amplitude and classifying each graph according to its topology. The result is an infinite set of coupled equations for the amplitudes of the field theory. Some approximate truncation of this set of equations (e.g. one which preserves $s$-channel unitarity) must be found. Since these equations are derived directly from the field theory, all of the integrals present in them will be four-dimensional. Note that despite the use of the Feynman diagrammatic series, Taylor's method is a non-perturbative technique, since it sums certain classes of graphs to all orders in the coupling constant. The great advantage of this method is that the resulting equations depend only on the topology of the allowed vertices of the field theory. Hence this is a particularly useful approach in hadronic field theories, where the best effective Lagrangian has not yet been derived from QCD.

When applied to the two-to-two amplitude the Taylor method yields

$$T^{(1)} = T^{(2)} + T^{(2)} d_1 d_2 T^{(1)}, \tag{1.1}$$

where the bracketed superscripts indicate the irreducibility of each amplitude. This is, of course, a Bethe-Salpeter equation (BSE) for the two-to-two amplitude, with $T^{(2)}$ as the "driving term" [9,10]. We note that if $T^{(2)}$ is the sum of *all* two-particle irreducible two-to-two Feynman diagrams, then this equation is exact. If the Taylor method is applied to the three-to-three amplitude, four-dimensional Faddeev-like equations may be derived [11].

In hadronic physics one problem which one might try to apply Eq. (1.1) to is nucleon-nucleon scattering. Before any calculations can be performed the nature of the kernel $T^{(2)}$ must be specified. One approach, pursued by Tjon and Fleischer, is to assume $T^{(2)}$ consists of a sum of one-boson-exchange Feynman diagrams. This approximation gives a reasonable fit to the nucleon-nucleon phase shifts [12,13]. However, for energies above the pion-production threshold, one needs to include the thresholds for all of the physically allowed channels. In particular, if a theory which gives consistent predictions for $NN$ scattering and pion production is to be developed, the full $NN\pi$ intermediate-state structure of the amplitude $T^{(2)}$ must be exposed. The Taylor method is well suited to this task, and attempts to use it in order to expose the $NN\pi$ intermediate-state structure of field-theoretic amplitudes, such as $T^{(2)}$, were first pursued by Avishai and Mizutani [14,15]. They used Taylor's method to classify all of the covariant perturbation theory diagrams which contribute to the $NN \to NN$, $NN \leftrightarrow NN\pi$ and $NN\pi \to NN\pi$ amplitudes and contain one explicit pion. The result was a set of Faddeev-like coupled four-dimensional integral equations for the processes

$$\left.\begin{array}{c} N+N \\ N+N+\pi \end{array}\right\} \leftrightarrow \left\{\begin{array}{c} N+N \\ N+N+\pi \end{array}\right. . \tag{1.2}$$

---

[*]Present address: Department of Physics, University of Maryland, College Park, Maryland, 20742, USA
[†]Email: phillips@quark.umd.edu, Iraj.Afnan@flinders.edu.au

These equations were formally equivalent to the three-dimensional $NN - \pi NN$ equations previously obtained by a number of groups using different techniques [14,16–22]. However, a crucial difference was that Avishai and Mizutani's equations contained four-dimensional, rather than three-dimensional, integrals. Similar four-dimensional equations, but for the $BB - \pi BB$ system, where the $B$ could be either a nucleon or a delta, were later obtained by Afnan and Blankleider [23]. By explicit construction, both Avishai and Mizutani's and Afnan and Blankleider's equations were covariant. Furthermore, unlike the three-dimensional $NN - \pi NN$ equations, they did not treat different time-orders of the same physical process differently. Consequently, these four-dimensional equations avoided the theoretical problems of the three-dimensional $NN - \pi NN$ equations which were pointed out by Sauer, Sawicki & Furui [24] and Jennings & Rinat [25,26]. (For a more detailed discussion of this point see [27,28].)

Unfortunately, these four-dimensional $NN - \pi NN$ equations were not correct, since they double-counted certain Feynman diagrams. This double-counting occurred due to a problem with the original Taylor method. Indeed, such double-counting will arise whenever one tries to derive coupled field-theoretic equations whose input is full sub-amplitudes for sub-system interactions. While a decomposition in terms of sub-system interactions is possible for any Feynman diagram, it is not necessarily unique. Therefore, when the perturbation series is reorganized in order to derive such coupled equations, some Feynman diagrams get incorrectly included more than once in the resummed series. (See [29] for a brief discussion and [8] for details.) To overcome this problem, we need a procedure for determining the conditions under which double-counting occurs, and a way to remove the offending diagrams from the equations. A "modified Taylor method", which gives a prescription for the subtractions necessary to remove this double-counting, was recently developed by us [8]. We then used this modified Taylor method to derive revised four-dimensional $NN - \pi NN$ equations which do not contain double-counting [27,29]. These equations are essentially equivalent to those obtained by Kvinikhidze and Blankleider using different means [30,31]. The resulting set of coupled four-dimensional two-fragment scattering equations obey $NN$ and $NN\pi$ unitarity, are free of double-counting, and are covariant, not only on-shell, but also off-shell in the manner dictated by the Feynman diagrammatic expansion.

Thus a set of equations which gives a complete and correct summation of the $NN$ and $NN\pi$ sectors of any hadronic field theory now exists. However, it remains to be seen whether this set of equations can be solved in order to yield predictions for experimental quantities. For a detailed program describing how such a solution might be achieved, see Section 7 of Ref. [27]. This paper reports on a preliminary study of the numerical solution of these equations. The equations are solved, not in a field theory of nucleons and pions, but in a scalar field theory. Consequently, details of spin and isospin are removed from the problem. In addition, there is no need to introduce subtractions or form factors to obtain convergent integrals. The simplified calculations described here imply that the numerical solution of the full four-dimensional $NN - \pi NN$ equations *is* feasible. The use of a scalar field theory also makes for a clear discussion of issues such as the usefulness of a coupled-channels formulation and the importance of double-counting removal in such an approach.

A field theory describing two types of spin-0 particles interacting via a Yukawa coupling is chosen. A contact interaction is also included, in order to ensure the stability of the classical vacuum. The Lagrangian density for this "$\phi^2\sigma$ field theory" is

$$\mathcal{L} = \frac{1}{2}[(\partial_\mu \phi)^2 - m^2\phi^2 + (\partial_\mu \sigma)^2 - \mu^2\sigma^2] - \frac{g}{2}\phi^2\sigma - \frac{\beta}{4}\phi^2\sigma^2. \tag{1.3}$$

Because the modified Taylor method is Lagrangian independent, the $NN - \pi NN$ equations for distinguishable particles derived in Ref. [27] apply to the $\phi^2\sigma$ field theory. The distinguishable-particle equations must be symmetrized, as the $\phi$'s are bosons. This leads to a set of nine two-fragment scattering equations for the processes

$$\left.\begin{array}{c} \phi+\phi \\ (\phi\sigma)+\phi \\ (\phi\phi)_d + \sigma \end{array}\right\} \longrightarrow \left\{\begin{array}{c} \phi+\phi \\ (\phi\sigma)+\phi \\ (\phi\phi)_d + \sigma \end{array}\right. , \tag{1.4}$$

where $(\phi\phi)_d$ is a bound state of the $\phi\phi$ system that plays the role of the deuteron. Below the threshold for $\sigma$ production these equations provide a way of calculating the $\phi - \phi$ amplitude which corresponds to including infinitely many diagrams in the kernel of a single Bethe-Salpeter equation for that amplitude. Above $\sigma$-production threshold these coupled equations allow us to calculate amplitudes for the reactions of Eq. (1.4) in a framework which—provided the input amplitudes have two-body unitarity—maintains $\phi\phi$ and $\phi\phi\sigma$ unitarity. Note that as in calculations using a single BSE, the kernel to these equations must be approximated in order to make them computationally tractable.

The inputs to these $\phi\phi - \sigma\phi\phi$ equations are the one-particle irreducible $\sigma - \phi$ amplitude $t^{(1)}_{\phi\sigma}$ and the $\phi\phi\sigma$ vertex function $f^{(1)}$. Because the $\phi\phi \to \phi\phi$ amplitude appears in the kernel of these equations they are a set of non-linear, or bootstrap, equations. Strictly speaking the complete non-linear set of four-dimensional equations should be solved for the amplitudes of interest. To snap the bootstrap, in this work we mimic the approach of previous three-dimensional $\pi N - \pi\pi N$ calculations [32] and use an input $\phi\phi$ amplitude which is constructed as a separable approximation to the "true" $\phi - \phi$ amplitude. The energy range used in the construction of this expansion must be $\mu$ below the energy at which the coupled equations are to be solved. Arguments from Ref. [27] show that we expect such an approach to yield a reasonable solution to the set of non-linear coupled equations. In this work we take the predictions of the ladder Bethe-Salpeter equation as the "true" $\phi - \phi$ amplitude. The consistency of this procedure may be judged by comparing the result of the coupled-channels calculation to the separable input amplitude.

In order to simplify matters as far as possible we assume that the amplitude $t^{(2)}_{\phi\sigma}$ is zero, i.e. that $\phi\sigma$ scattering is dominated by the pole diagram. The implications of imposing this restriction are discussed in Sec. II. This assumption implies that the vertex function for $\sigma$ absorption on a single $\phi$ is given by

$$f^{(1)}(p', p, k) = f^{(2)}(p', p, k) = -ig(2\pi)^4 \delta^{(4)}(p' - p - k), \tag{1.5}$$

where the Feynman rules of the field theory have been used in order to evaluate $f^{(2)}$.

If the input amplitudes are chosen in this way then the $\phi\phi - \sigma\phi\phi$ equations become a set of four coupled equations describing the processes in (1.4) which do not involve the $(\phi\sigma)$ "isobar". In this paper we solve these equations for $\phi\phi$ scattering up to the $\sigma$-"deuteron" threshold and compare the results with those from the BSE. In so doing we lay the foundation for studies of the covariant, four-dimensional, $NN - \pi NN$ equations derived in Ref. [27].

In Sec. II we consider the full $\phi\phi - \sigma\phi\phi$ equations and discuss the consequences of the approximations made to get the simplest form of the equations which includes the coupling to the $\sigma(\phi\phi)_d$ channel. In the process we see how the double-counting problem arises at each level of approximation, and how the content of the kernel compares with that of the BSE. We also examine the consequences of taking the ladder approximation for the $\phi - \phi$ amplitude in the kernel of the coupled integral equations.

Since the input to the equations derived in Sec. II is the $\phi - \phi$ amplitude, we proceed in Sec. III to discuss the numerical solution of the Bethe-Salpeter equation in the ladder approximation. We demonstrate that, in the case when the interaction supports a bound system with a binding energy comparable to that of the deuteron, a valid separable approximation to the $\phi - \phi$ amplitude can be written.

In Sec. IV we examine the solution of the coupled integral equations for the reactions in Eq. (1.4). At this stage only the solution of these equations below the threshold for $\sigma$ production is explored. We then compare our results with those from the BSE when the kernel is taken to be either the ladder, or the ladder plus crossed-box $\sigma$ exchange. The effect of vertex dressing in the BSE is also investigated. In particular, it is found that vertex dressing must be included in the Bethe-Salpeter kernel in order to maintain consistency with the coupled-channels results. Finally, in Sec. V we give a summary with some concluding remarks about future work.

## II. THE $\phi\phi - \sigma\phi\phi$ EQUATIONS

There are two central questions in the analysis of the $\phi\phi - \sigma\phi\phi$ equations. Firstly, what additional features do these equations have as compared to BSE formulations? In particular, how does the physics content of these equations change as different approximations are made for the kernel? Secondly, what numerical problems are encountered in solving these equations? In this section we address the first of these questions, by illustrating the origin of the double-counting problems and showing how subtractions remove them. The $NN - \pi NN$ equations reported in Ref. [27,29], which are basically the same as the equations of Kvinikhidze and Blankleider [30,31], are also valid for the Lagrangian in Eq. (1.3) since the derivation of the equations does not depend on the choice of Lagrangian. As a result, the $\phi\phi - \sigma\phi\phi$ equations for distinguishable $\phi$'s are of the form

$$\begin{aligned} T_{\phi\phi} &= \bar{V} \left( 1 + d_1 d_2 \, T_{\phi\phi} \right) \\ &\quad + \sum_{j\alpha} f^{(1)}(j) \, \bar{\delta}_{j\alpha} \, d_{\bar{j}} \, d_\sigma \, \tilde{t}^{(1)\dagger}(\alpha) \, d_\alpha^{-1} \, d_1 d_2 d_\sigma \, T_{\alpha\phi} \\ T_{\lambda\phi} &= \sum_j \bar{\delta}_{\lambda j} \, f^{(1)\dagger}(j) \, d_j^{-1} \left( 1 + d_1 d_2 \, T_{\phi\phi} \right) \end{aligned} \tag{2.1}$$

$$+ \sum_\alpha \bar{\delta}_{\lambda\alpha} \, t^{(1)}(\alpha) \, d_\alpha^{-1} \, d_1 d_2 d_\sigma \, T_{\alpha\phi}$$
$$- \sum_{ij} v^x(i) \, d_\sigma \, \bar{\delta}_{ij} \, f^{(1)\dagger}(\bar{j}) \, (1 + d_1 d_2 T_{\phi\phi}) \, , \tag{2.2}$$

where we have taken particles one and two to be the $\phi$ particles, while particle three is the $\sigma$. Here, $\bar{\delta}_{ij} = 1 - \delta_{ij}$ and $\bar{j} = 2$ for $j = 1$, while $\bar{j} = 1$ for $j = 2$. The roman indices run from one to two, while the greek indices run from one to three. These equations differ from the corresponding equations of Avishai and Mizutani [15] and Afnan and Blankleider [23] in that the $\phi-\phi$ potential $\bar{V}$ and the one-particle irreducible $2 \to 2$ amplitude $\tilde{t}^{(1)\dagger}(\alpha)$ in Eq. (2.1) have subtractions to remove double counting in the $\phi - \phi$ amplitude $T_{\phi\phi}$, while the last term in Eq. (2.2) is a subtraction to avoid double counting in the $\phi\phi \to \sigma\phi\phi$ amplitude $T_{\lambda\phi}$. $f^{(1)}(i)$ is the one-particle irreducible $\phi \leftarrow \phi\sigma$ amplitude with particle $i$ as spectator, while $t^{(1)}(\alpha)$ is the one-particle irreducible $2 \to 2$ amplitude with particle $\alpha$ as spectator. The single-particle Feynman propagator for particle $\alpha$ is $d_\alpha$. The $\sigma - \phi$ potential $v^x$ in Eq. (2.2) corresponds to the $u$-channel $\phi$-pole diagram, i.e.

$$v^x(i) = f^{(1)}(i) \, d_{\bar{i}} \, f^{(1)\dagger}(i) \, . \tag{2.3}$$

The two-body amplitude $\tilde{t}^{(1)}$ is chosen such that the subtraction is present only in the $\sigma - \phi$ amplitudes (see Ref [27]), with the subtraction basically removing any initial $u$-channel pole, i.e.

$$\tilde{t}^{(1)}_{\phi\phi} = t^{(1)}_{\phi\phi}; \quad \tilde{t}^{(1)}_{\sigma\phi} = \left( t^{(1)}_{\sigma\phi} d_\phi d_\sigma + 1 \right) v^R_{\sigma\phi} \, ; \tag{2.4}$$

$$v^R_{\sigma\phi} = t^{(2)}_{\sigma\phi} - v^x_{\sigma\phi} \, . \tag{2.5}$$

In Eq. (2.1) the $\phi - \phi$ potential $\bar{V}$ is given by

$$\bar{V} = V^*_{OSE} - D_1 - D_2 - D_\sigma - X - B \, , \tag{2.6}$$

where $V^*_{OSE}$ is the one-$\sigma$-exchange potential, i.e.,

$$V^*_{OSE} = f^{(1)*}(1) \, d_\sigma \, f^{(1)*\dagger}(2) \, , \tag{2.7}$$

with the modified one-particle irreducible $\sigma\phi\phi$ form factor, $f^{(1)*}$, having dressing with the $u$-channel pole diagram subtracted, i.e.

$$f^{(1)*} = f^{(2)} \left( 1 + d_\sigma d_\phi \tilde{t}^{(1)}_{\sigma\phi} \right) \, . \tag{2.8}$$

The subtractions $D_2$, $D_\sigma$, $X$, and $B$ are illustrated in Fig. 1, and are given explicitly in Ref [27]. $D_1$ is merely $D_2$ with the two nucleons interchanged. All sub-amplitudes in Fig. 1 are one-particle irreducible in the $s$-channel.

Since we are ultimately interested in the $NN - \pi NN$ equations, the first approximation we consider is that the two-particle irreducible $\sigma - \phi$ amplitude is just the $u$-channel pole diagram, i.e. $t^{(2)}_{\sigma\phi} = v^x_{\sigma\phi}$. In this approximation the subtractions $D_1$ and $D_2$ in (2.6) become unnecessary. All of the other double-counting problems are still present, and must be removed by the explicit subtraction terms. Since $v^R_{\sigma\phi} = 0$, the form factors in the $\sigma$-exchange potential are two-particle irreducible, while the $u$-channel subtracted $\sigma - \phi$ amplitude is zero, i.e.

$$f^{(1)*} = f^{(2)} \qquad \tilde{t}^{(1)}_{\sigma\phi} = 0 = \tilde{t}^{(1)\dagger}_{\sigma\phi} \, . \tag{2.9}$$

We could now approximate both the $\sigma - \phi$ and the $\phi - \phi$ amplitudes by separable expansions, which, after partial-wave expansion, would reduce the coupled-channels problem to a set of coupled integral equations in two dimensions. However, since this is the first numerical work on this problem, we have chosen to concentrate on the problem of the separable expansion for the $\phi - \phi$ amplitude, and the procedure for linearization of the coupled non-linear integral equations. To this end we have chosen $t^{(2)}_{\sigma\phi} = 0$. Such a choice implies that all one-particle irreducible form factors in Eqs. (2.1) and (2.2) be replaced by the corresponding two-particle irreducible form factor, and the one-particle irreducible $\sigma - \phi$ amplitude $t^{(1)}_{\sigma\phi}$ be set to zero, i.e., $f^{(1)} \to f^{(2)}$ and $t^{(1)} \to 0$. This reduces our coupled integral equations to two coupled equations of the form

$$T_{\phi\phi} = \bar{V} \left( 1 + d_1 d_2 \, T_{\phi\phi} \right)$$
$$+ \sum_j f^{(2)}(j) \, d_{\bar{j}} d_\sigma \, t^{(1)}_{\phi\phi} \, d_1 d_2 \, T_{d\phi} \tag{2.10}$$

$$T_{d\phi} = \sum_j f^{(2)\dagger}(j) \, d_j^{-1} \left( 1 + d_1 d_2 \, T_{\phi\phi} \right) . \tag{2.11}$$

In writing these equations we have made use of the notation $T_{d\phi} \equiv T_{3\phi}$. In these equations the correction required to remove the double-counting problem is restricted to the $\phi - \phi$ potential $\bar{V}$. This is because all double counting due to $u$-channel poles in the $\sigma\phi$ t-matrix has been, by definition, removed from the problem.

At this stage we should recall that the one-particle irreducible $\phi - \phi$ amplitude, $t^{(1)}_{\phi\phi}$, is equal to the full amplitude, $T_{\phi\phi}$, which is part of the solution to the coupled equations, and so we have a non-linear set of integral equations. In the time-ordered, or non-relativistic, version of these equations, the energy variable in $t^{(1)}_{\phi\phi}$ is lower than the energy in the amplitude $T_{\phi\phi}$ by at least the mass of the $\sigma$. This allows for a bootstrap procedure, in which the lower energy amplitude is used to obtain the amplitude at higher energies—as illustrated in work on the $\pi N - \pi\pi N$ equations above the threshold for pion production [32]. A similar situation prevails in the covariant framework—but only approximately (see [27,28] for details). In the present analysis we will follow this procedure for removing the non-linearity. To simplify the resulting bootstrap problem we will assume that $t^{(1)}_{\phi\phi}$ is a solution of the BSE in the ladder approximation. This final approximation removes the bootstrap problem, and our coupled-channel equations still satisfy two and three-body unitarity. Furthermore, we can test the validity of this solution to the bootstrap problem by comparing the phase shifts corresponding to $t^{(1)}_{\phi\phi}$ with the phase shifts resulting from the solution of the coupled-channels problem.

Before we turn to further discussion of the input to the coupled-channel equations, we should point out that the above two coupled equations can be recast into a single equation for the $\phi - \phi$ amplitude

$$T_{\phi\phi} = V_{eff} \left( 1 + d_1 d_2 \, T_{\phi\phi} \right) , \tag{2.12}$$

with an effective interaction of the form

$$V_{eff} = \bar{V} + \sum_{ij} f^{(2)}(i) \, d_{\bar{i}} d_\sigma \, t^{(1)}_{\phi\phi} \, d_{\bar{j}} \, f^{(2)\dagger}(j) . \tag{2.13}$$

Rewriting the coupled-channels problem in terms of a single channel allows us to see how the double counting is removed by subtractions in $\bar{V}$. It also helps in discussions of the difference between the coupled-channel equations and the ladder, or ladder plus crossed-box, BSE.

In the event that $t^{(1)}_{\phi\phi}$ is chosen to be the solution of the ladder BSE, then the second term on the right hand side (r.h.s.) of Eq. (2.13) has two terms that correspond to crossed-box $\sigma$-exchange. This double counting is removed by taking

$$\bar{V} = V^*_{OSE} - B , \tag{2.14}$$

where $B$ is the crossed-box digram (see Fig. 1). In addition, when $i = j$, the second term on the r.h.s. of Eq. (2.13) has diagrams that contribute to vertex dressing of the $\sigma$-exchange potential $V^*_{OSE}$ (diagrams $(a)$ and $(b)$ of Fig. 2). However, a careful examination of the series resulting from this term reveals that there are no contributions from diagrams in which both vertices are being dressed at the same time. This is a result of our elimination of the non-linearity in the coupled integral equations. Finally, expanding the $i \neq j$ piece of the second term on the r.h.s. of Eq. (2.13) shows that some of the crossed-ladder diagrams required to get the correct one-body limit for the BSE [33] are included in the solution of Eq. (2.12). Indeed, the solution of the non-linear equation for $T^{(1)}$ will include more (but not all) of the crossed-ladder diagrams necessary for the correct one-body limit.

In the event that the amplitude from the solution of Eq. (2.12) is not consistent with the solution of the ladder BSE, we could take the solution of Eq. (2.12) and substitute that result in Eq. (2.13). This in turn will give us a new effective potential that can be used in the BSE, Eq. (2.12). However, in that case more diagrams are generated by the second term on the r.h.s. of Eq. (2.13), and therefore further subtractions are needed. . Consequently, the potential $\bar{V}$ is given by

$$\bar{V} = V^*_{OSE} - B - D_\sigma - X . \tag{2.15}$$

This demonstrates very clearly that when the input to the equations is approximated we need to be careful in calculating the correction for the double-counting problem. We return to this point in Sec. IV D when we consider including vertex dressing in the BSE.

## III. THE BETHE-SALPETER EQUATION

In most numerical analyses of the $NN - \pi NN$ equations the input two-body amplitudes were parameterized to fit the experimental data, and no regard was given to the consistency between the input $N - N$ amplitude and the $N - N$ amplitude resulting from the solution of the coupled-channels problem. In the present analysis there is no experimental data for the $\phi - \phi$ system, and we would like to specifically maintain some consistency between the input $\phi - \phi$ amplitude and that resulting from the solution of the coupled-channels problem. Therefore, as we have detailed in the previous section, we will consider the coupled $\phi\phi - \sigma\phi\phi$ equations with the only input amplitude being the $\phi - \phi$ amplitude in the ladder approximation. In this way the problem of the non-linearity of the coupled-channel equations is avoided. After partial-wave expansion, the resultant coupled-channels problem is a set of coupled integral equations in four dimensions. To reduce the dimensionality of our equations we follow the procedure previously implemented for the $\pi N - \pi\pi N$ equations [32], and construct a separable approximation to the solution of the ladder BSE. In this way we reduce the dimensionality of the coupled integral equations from four to two, at the possible cost of increasing the number of coupled integral equations.

Therefore in this section we explain various aspects of the solution of the Bethe-Salpeter equation and the construction of our two-body input. These include techniques for: the solution of the ladder BSE; calculation of bound-state masses; construction of a separable approximation to the ladder amplitude; and inclusion of effects beyond the ladder approximation. While all of the techniques described here are explained elsewhere [34–38], we give an outline of the methods for two reasons. Firstly, our solution procedures for the coupled-channel equations are extensions of those described here. Secondly, explaining the methods establishes some of the notation used in Section IV. Further details on the work of this section may be found in Ref. [28].

### A. The ladder approximation

The BSE, Eq. (1.1), is an equation for the invariant two-particle scattering amplitude in a field theory. Of course, in practice it is impossible to construct the sum of all graphs contributing to $T^{(2)}$ and solve the resulting integral equation for $T^{(1)}$. Therefore, most work in the past has focused on the so-called "ladder approximation", in which $T^{(2)}$ is taken to consist solely of one-particle exchange.

Consider the $\phi^2 \sigma$ field theory. Define:

$$T^{(1)}(p'_1, p'_2; p_1, p_2) = -i(2\pi)^4 \delta^{(4)}(p'_1 + p'_2 - p_1 - p_2) T(q', q; P) . \tag{3.1}$$

If the total momentum is $P = p'_1 + p'_2 = p_1 + p_2$, then in the kinematics specified by Fig. 3 the BSE in ladder approximation is

$$T(q', q; P) = V(q', q) + \frac{i}{(2\pi)^4} \int d^4 q'' \, V(q', q'') \, d_\phi(q''_+) \, d_\phi(q''_-) \, T(q'', q; P) . \tag{3.2}$$

Here $q''_\pm = \frac{P}{2} \pm q''$, while the potential due to $\sigma$-exchange and the $\phi$ propagator are defined by

$$V(q', q) = \frac{g^2}{(q' - q)^2 - \mu^2} , \quad d_\phi(p) = \frac{1}{p^2 - m^2} . \tag{3.3}$$

Throughout this paper the masses $m$ and $\mu$ are regarded as having an infinitesimal negative imaginary part. This defines the way poles should be negotiated, unless otherwise stated in the text.

The BSE may also be used to extract information about bound-state properties. It is known that if a bound state of the two $\phi$ particles with mass $M$ exists then the invariant scattering amplitude may be decomposed into a pole and non-pole piece [39]. Inserting this decomposition into Eq. (3.2) and taking the residue at the pole yields

$$\Gamma(q'; M^2) = \frac{i}{(2\pi)^4} \int d^4 q'' \, V(q', q'') \, \left[d_\phi(q''_+) d_\phi(q''_-)\right]_{P^2 = M^2} \Gamma(q''; M^2) , \tag{3.4}$$

where $\Gamma$ is the bound-state vertex function. Thus the bound-state solutions of the ladder BSE may be found by searching for those values of $P^2$ for which the kernel of the BSE in the ladder approximation has an eigenvalue of one.

The BSE in the ladder approximation is a singular integral equation and as it stands is not amenable to numerical solution. One possible way of dealing with the singular nature of the equation is to perform a Wick rotation [40], which involves an analytic continuation of the two variables $q'_0$ and $q''_0$ in Eq. (3.2) to the imaginary axis. In the center-of-mass frame, where $P = (\sqrt{s}, \vec{0})$, this analytic continuation requires that we take into consideration the singularities of both the kernel and the Bethe-Salpeter amplitude. The singularities encountered depend on the total energy available, $\sqrt{s}$. As explained by Taylor and Paganamenta [41], this straightforward analytic continuation to the imaginary axis is valid for $0 \leq \sqrt{s} \leq 2m + 2\mu$. Above $\sqrt{s} = 2m + 2\mu$ the self-consistently-generated $\phi\sigma$ production threshold cuts in the amplitude $T$, which begin at $q_0'' = \pm(\frac{\sqrt{s}}{2} - m - \mu)$, enter the first and third quadrants, thus pinching the $q_0''$ integration contour, and preventing a simple analytic continuation of Eq. (3.2). Note that for $2m \leq \sqrt{s} \leq 2m + 2\mu$ poles of the $\phi$ propagators may enter the first and third quadrants of the $q_0''$ plane. If this occurs then the residue at these poles must be included in the Wick-rotated equations, in order to ensure the correct analytic continuation.

To reduce the dimensionality of Eq. (3.2) from four to two, and so be able to turn the resultant equation into an algebraic equation, we need to partial-wave expand both the amplitude $T$ and the one-$\sigma$-exchange potential. The partial-wave expansion

$$A(q_\mu', q_\mu; s) = (-8\pi^3) \sum_\ell \frac{(2\ell+1)}{2q'q} A_\ell(q_0', q', q_0, q; s) P_\ell(\cos\theta) , \qquad (3.5)$$

where $A$ is either the one-$\sigma$-exchange potential $V$ or the amplitude $T$, and $\cos\theta = \hat{q}\cdot\hat{q}'$, is used. In the present analysis we take $\hbar = c = 1$, and the unit of energy is chosen such that $m = 1$. From this point all four-vectors have a subscript $\mu$, so as to distinguish them from the magnitude of three-vectors. The partial-wave expansion converts Eq. (3.2) to

$$T_\ell(q_0', q', q_0, q; s) = V_\ell(q_0', q', q_0, q) - i \int_{-\infty}^{\infty} dq_0'' \int_0^{\infty} dq'' \, V_\ell(q_0', q', q_0'', q'') \, G(q_0'', q''; s) \, T_\ell(q_0'', q'', q_0, q; s) , \qquad (3.6)$$

where:

$$G(q_0'', q''; s) = \frac{1}{(\frac{\sqrt{s}}{2} + q_0'')^2 - E_\phi(q'')^2} \frac{1}{(\frac{\sqrt{s}}{2} - q_0'')^2 - E_\phi(q'')^2} , \qquad (3.7)$$

$$V_\ell(q_0', q', q_0, q) = \frac{2\lambda}{\pi} Q_\ell\left(\frac{q'^2 + q^2 + \mu^2 - (q_0' - q_0)^2}{2q'q}\right) , \qquad (3.8)$$

$$E_\phi(q'') = \sqrt{q''^2 + m^2} - i\epsilon ; \qquad \lambda = \frac{g^2}{16\pi^2} ; \qquad (3.9)$$

and $Q_\ell$ is a Legendre function of the second kind.

The on-shell partial-wave amplitude $T_\ell^{on}$ can now be written in terms of the phase shift $\delta_\ell$ as

$$T_\ell^{on}(s) = T_\ell(0, \bar{q}, 0, \bar{q}; s) = \frac{2\bar{q}\sqrt{s}}{\pi^2} e^{i\delta_\ell(s)} \sin\delta_\ell(s) , \qquad (3.10)$$

where the on-shell relative momentum is given by $\bar{q} = \sqrt{\frac{s}{4} - m^2}$.

Next we perform a Kowalski-Noyes (KN) [42–44] reduction, in order to remove the two-body unitarity cut, i.e. we write

$$T_\ell(q_0'', q'', 0, \bar{q}; s) = f_\ell(q_0'', q''; s) T_\ell^{on}(s) , \qquad (3.11)$$

with the KN half-off-shell function being one at the on-shell point, i.e. $f_\ell(0, \bar{q}; s) = 1$. This is followed by a Wick rotation. Since the two $\phi$ particles have the same mass, we change the $q_0''$ integration to one from zero to infinity, and so obtain the coupled equations derived by Levine et al. for $f_\ell(iq_0', q'; s)$ and $g_\ell(q'; s)$ [34,45]. The auxiliary equation for $g_\ell(q'; s)$ must be written because of the presence of $\phi$-propagator poles in the first and third quadrants of the $q_0''$ plane. $g_\ell(q; s)$ is defined by

$$g_\ell(q; s) = f_\ell(\bar{w}(q), q; s) , \qquad \bar{w}(q) = \frac{\sqrt{s}}{2} - E_\phi(q) . \qquad (3.12)$$

The equations for $f_\ell$ and $g_\ell$ can be solved by discretizing the integrals and using direct matrix inversion. We note that there is a term $V_\ell(\bar{w}(q'), q', -\bar{w}(q''), q'')$ in the kernel of the integral equations. For $2m + \mu < \sqrt{s} < 2m + 2\mu$ this term may have a logarithmic branch-point in the region of integration. We found that the equations could still be solved accurately in this energy regime, provided enough quadratures were used for the discretization.

To calculate phase shifts, the equation

$$T_\ell^{on}(s) = \frac{V_\ell^{on}(s)}{1 + i \int_{-\infty}^{\infty} dq_0'' \int_0^{\infty} dq'' \, V_\ell(0, \bar{q}, q_0'', q'') \, G(q_0'', q''; s) \, f_\ell(q_0'', q''; s)} \quad . \tag{3.13}$$

is used to obtain $T_\ell^{on}(s)$. In order to deal with the singularities in the integrand the integral is written as [46]

$$i \int_{-\infty}^{\infty} dq_0'' \int_0^{\infty} dq'' \, V_\ell(0, \bar{q}, q_0'', q'') \, G(q_0'', q''; s) \, f_\ell(q_0'', q''; s) = 2a(s) + b(s) \, , \tag{3.14}$$

where the factors $a(s)$ and $b(s)$ are given by

$$a(s) = -\int_0^{\infty} dq_0'' \int_0^{\infty} dq'' \, G(iq_0'', q''; s) \, (V_\ell(0, \bar{q}, iq_0'', q'') \, f_\ell(iq_0'', q''; s) - V_\ell^{on}(s))$$
$$+ \, \pi \int_0^{\bar{q}} dq'' \, \frac{1}{2\sqrt{s}\bar{w}(q'')E_\phi(q'')} \, (V_\ell(0, \bar{q}, \bar{w}(q''), q'') \, g_\ell(q''; s) - V_\ell^{on}(s)) \, , \tag{3.15}$$

$$b(s) = V_\ell^{on}(s) \int_{-\infty}^{\infty} dq_0'' \int_0^{\infty} dq'' \, G(q_0'', q''; s) \, . \tag{3.16}$$

To test the accuracy of our numerical procedure, we compared our results with those reported by Levine *et al.* [34], and found agreement. Varying the quadrature distribution then shows that with 14 and 22 Gauss-Legendre quadratures for the $q_0''$ and $q''$ integration in the Wick rotated integral, and 60 in the auxiliary residue $q''$ integral, the real part of the phase shift is accurate to three significant figures, while the imaginary part is accurate to two significant figures. The large number of quadratures is necessary in the auxiliary integral because of the presence of logarithmic branch-points in the kernel if $\sqrt{s} > 2m + \mu$.

Since we hope to have in the $\phi\phi - \sigma\phi\phi$ system a simple model for the $NN - \pi NN$ system, we have chosen to construct the input $\phi - \phi$ amplitude to have a bound state with a binding comparable to that of the deuteron. To achieve this we consider the homogeneous BSE in ladder approximation after partial-wave expansion,

$$\Gamma_\ell(q_0', q'; s) = -i \int_{-\infty}^{\infty} dq_0'' \int_0^{\infty} dq'' \, V_\ell(q_0', q', q_0'', q'') \, G(q_0'', q''; s) \, \Gamma_\ell(q_0'', q''; s) \, , \tag{3.17}$$

where $\Gamma_\ell(q_0', q'; s)$ is the vertex function for relative four-momentum $q_\mu'$, with its angular dependence given by $P_\ell(\hat{q}' \cdot \hat{z})$. This may be rewritten as an eigenvalue equation with a symmetric kernel, $\tilde{K}_\ell$. Note that because $\tilde{K}_\ell$ is linear in the coupling $\lambda$ it is not actually necessary to calculate the eigenvalue at a number of energies. Instead we merely write $\tilde{K}_\ell = \lambda \tilde{K}_\ell'$. The largest eigenvalue of $\tilde{K}_\ell'(s)$ is then the inverse of the coupling $\lambda$ required to produce a bound state of energy $s$ with angular momentum $\ell$. Using this technique we obtained, for the case $\mu = 0.15$, the plot of bound-state energy against $\lambda$ shown in Fig. 4. If a result accurate to three decimal places is desired the $q_0''$ and $q''$ integrations require 12 and 22 quadratures respectively.

The coupling $\lambda$ is chosen so that the $S$-wave $\phi - \phi$ bound state has the same binding energy as the deuteron. In our units $m_d^2 = 3.9905$. The value $\lambda = 0.131$ yields a bound state of this mass. The value of the "deuteron" binding energy is only accurate to 5%; however, for our purposes this is more than adequate.

### B. Separable approximation

Having determined the Bethe-Salpeter amplitude in the ladder approximation, we now need to get a separable representation of this amplitude, so that we can solve the $\phi\phi - \sigma\phi\phi$ equations numerically. Ideally, a separable expansion of the one-$\sigma$-exchange amplitude in terms of the eigenstates of the kernel, or the half-off-mass shell solution to the BSE in the ladder approximation [47], should be used. However, as a first attempt at a solution of the $\phi\phi - \sigma\phi\phi$ equations, we consider the phase shifts from the ladder BSE as data, and fit this data with a rank-one separable potential.

In a given partial wave, we can write a covariant version of a rank-one separable potential, $V_\ell$, as [36]

$$V_\ell(q_0', q', q_0, q) = -\frac{1}{4\pi^3} q' \chi_\ell(q_0', q') \xi_\ell \, \chi_\ell(q_0, q) q \ . \tag{3.18}$$

The factors of $q$ and $q'$ are introduced into the potential to maintain consistency with our BSE, Eq. (3.6). Since we will be restricting our analysis to $S$-wave scattering, we have chosen the form factor $\chi_\ell$ for $\ell = 0$ to be a function of the four-momentum square,

$$\chi_0(q_0, q) = \frac{1}{q_0^2 - q^2 - \beta^2} \ . \tag{3.19}$$

The solution of the the BSE for this potential is also separable, being of the form

$$T_\ell(q_0', q', q_0, q; s) = -\frac{1}{4\pi^3} q' \, \chi_\ell(q_0', q') \, \tau_\ell(s) \, \chi_\ell(q_0, q) \, q \ , \tag{3.20}$$

where:

$$\tau_\ell^{-1}(s) = \frac{1}{\xi_\ell} - \int_{-\infty}^{\infty} dq_0'' \int_0^{\infty} dq'' [q'' \chi_\ell(q_0'', q'')]^2 \, G(q_0'', q''; s) \ . \tag{3.21}$$

Note that, assuming the existence of only one bound state in the channel with angular momentum $\ell$, $\tau_\ell$ may be rewritten

$$\tau_\ell(s) = \frac{S_\ell(s)}{s - m_d^2} \ , \tag{3.22}$$

where $m_d^2$ is the bound-state mass, to be determined by demanding that $\tau_\ell(s)$ has a pole at $s = m_d^2$, and $S_\ell(s)$ is the residue of $\tau_\ell(s)$ at that point. In Eq. (3.21) the $q_0''$ integration is performed analytically using the residue theorem [48]. For the scattering case, to remove the pinch at $q_0'' = 0$, $q'' = \bar{q}$, we rotate the contour of integration into the third quadrant of the $q''$-plane. An angle of rotation of $\frac{\pi}{8}$ and 32 quadratures for the $q''$ integration gives sufficient accuracy to fit the phase shifts resulting from the solution of the ladder BSE.

The parameter $\xi_0$ is adjusted to guarantee that the binding energy is identical to that predicted by the BSE equation in the ladder approximation with $\mu = 0.15$ and $\lambda = 0.13$, while the parameter $\beta$ is adjusted to get a best fit to the phase shifts below the production threshold. The final values for the two parameters in our choice of units are: $\xi_0 = -15.33$ and $\beta = 0.47386$. Despite our use of only two parameters, the fit to the 'data' is very good, as demonstrated in Fig. 5.

### C. Beyond the ladder approximation

In this section we consider two improvements to the standard ladder approximation, which will be used when we are comparing single-BSE calculations with results from the coupled-channels calculation. Firstly, we examine the inclusion of $\phi$ propagator dressing, which is necessary if $\phi\phi\sigma$ unitarity is not to be violated. Secondly, we explain how to include crossed-box $\sigma$-exchange in the kernel of the BSE.

The first improvement is to include the lowest-order dressing for the $\phi$ propagators. This effect was included in the derivation of the $\phi\phi - \sigma\phi\phi$ equations by implicitly assuming all propagators were dressed. In addition, the dressing of the propagators gives a contribution to three-body unitarity that complements the contribution of one-$\sigma$ exchange. Indeed, Levine *et al.* [34] have shown that such propagator dressing must be included in order to guarantee an inelasticity less than one above the threshold for $\sigma$ production. They found that for $\mu = m = 1$, and sufficiently large coupling, the dressing of the $\phi$ propagator plays an important role.

The inclusion of the minimal dressing in the $\phi$ propagator takes the form (see Fig. 6)

$$d(p^2) = \frac{1}{p^2 - m^2 - \Sigma(p^2) + i\epsilon} \ , \tag{3.23}$$

where to lowest order in the coupling

$$\Sigma(p^2) = g^2 \int \frac{d^4q}{(2\pi)^4} \frac{1}{(p-q)^2 - \mu^2} \frac{1}{q^2 - m^2} \ . \tag{3.24}$$

Note that we have not included the full one-loop self-energy term here, since in order to satisfy $\phi\phi\sigma$ unitarity it is only necessary to use bare propagators in the expression for $\Sigma$. We now write a dispersion relation for this propagator. Since we wish to examine the propagator after mass and coupling-constant renormalization this is a twice-subtracted dispersion relation. Algebraic manipulations á la Saenger [49] and Janus [50] yield:

$$d(p) = \frac{1}{p^2 - m^2 + i\epsilon} \frac{1}{1 + (p^2 - m^2)A(p^2)}, \tag{3.25}$$

with:

$$A(p^2) = \lambda \int_{(m+\mu)^2}^{\infty} ds' \frac{\sqrt{[s' - (\mu+m)^2][s' - (\mu-m)^2]}}{s'(s' - m^2)^2(s' - p^2 - i\epsilon)}. \tag{3.26}$$

The integral (3.26) may be evaluated numerically. We find that 18 quadratures are more than sufficient to ensure that it is accurately calculated.

Previously, Levine *et al.* [34] have shown that for $m = \mu = 1$ and a large coupling constant, the effect of the dressing on the phase shifts is substantial. Although we agree with their numerical results, we find that for $m = 1$ and $\mu = 0.15$, which correspond to the mass of the nucleon and pion, and a coupling strength that gives a binding energy for the $\phi - \phi$ state that is comparable to the deuteron binding, the effect of the dressing on the phase shifts is small. This is particularly the case for energies below the $\sigma$-production threshold, as demonstrated in Fig. 7, where we present the real part of the phase shifts. In fact, with dressed propagators a coupling of $\lambda = 0.13$ yields a bound-state of mass $m_d^2$—a very small change from the result for undressed propagators.

Despite this dressing having very little effect on the bound-state position and the real part of the phase shifts, it does have a substantial impact on the imaginary part of the phase shifts, as seen in Fig. 8. However, the imaginary part of the phase shifts is still small. This result is consistent with observations in $N - N$ scattering above pion-production threshold, where the contribution to pion production is dominated by the production of a $\Delta$ which then decays by pion emission, and the equivalent mechanism via the nucleon is comparatively small.

A more interesting addition to the ladder approximation is the inclusion of the crossed-box $\sigma$-exchange diagram in the potential. In particular, the coupled-channel equations with the ladder approximation for the input $\phi - \phi$ amplitude include such crossed diagrams, as pointed out in Sec. II. Furthermore, the inclusion of the crossed-box graph in the kernel partially corrects the failure of the ladder BSE in the one-body limit. If the BSE is written for two particles with unequal masses, then when the mass of one particle is taken to infinity we expect to recover an equation for the other, lighter, particle which represents it moving in the static potential generated by the heavy particle. If the ladder BSE is used this does not occur. This one-body limit is restored on including all ladders and crossed-ladders in the perturbation series for the amplitude [33]. Although the inclusion of the crossed diagram in the potential is not equivalent to including all crossed-ladder diagrams, the fact that our coupling constant is small implies that including this one extra diagram in the kernel will give us an indication of the contribution of higher-order crossed-ladder graphs.

Adding the crossed-box diagram (diagram $B$ of Fig. 1) to the kernel introduces no analytic structure into the problem beyond that discussed above, and therefore the Wick rotation may proceed exactly as for the ladder BSE. Indeed, the only change which needs to be made to our discussion of the ladder BSE is that $V_\ell$ now includes a piece from the crossed-box diagram.

In the case of the $\phi^2\sigma$ field theory the value of the crossed-box diagram may be calculated using the usual Feynman parameterizations [39], and the result is an amplitude

$$X(q'_\mu, q_\mu; s) = \int_0^1 d\alpha \int_0^{1-\alpha} d\beta \int_0^{1-\alpha-\beta} d\gamma \frac{16\pi^2 \lambda^2}{D^2(\alpha, \beta, \gamma | q'_\mu, q_\mu, s)}, \tag{3.27}$$

where:

$$\begin{aligned} D(\alpha, \beta, \gamma | q'_\mu, q_\mu; s) &= \alpha\gamma s_{11} + \beta(1-\alpha-\beta-\gamma)s_{22} \\ &\quad + \alpha(1-\alpha-\beta-\gamma)s_{33} + \beta\gamma s_{44} \\ &\quad + \alpha\beta u + \gamma(1-\alpha-\beta-\gamma)t \\ &\quad - (\alpha+\beta)m^2 - (1-\alpha-\beta)\mu^2, \end{aligned} \tag{3.28}$$

with:

$$s_{11} = (\frac{\sqrt{s}}{2} + q_0)^2 - q^2, s_{22} = (\frac{\sqrt{s}}{2} - q_0)^2 - q^2, \tag{3.29}$$

$$s_{33} = (\frac{\sqrt{s}}{2} + q'_0)^2 - q'^2, s_{44} = (\frac{\sqrt{s}}{2} - q'_0)^2 - q'^2, \tag{3.30}$$

$$t = (q_0 - q'_0)^2 - q^2 - q'^2 + 2qq'\hat{q} \cdot \hat{q}', \tag{3.31}$$

$$u = (q_0 + q'_0)^2 - q^2 - q'^2 - 2qq'\hat{q} \cdot \hat{q}'. \tag{3.32}$$

The amplitude $X$ can thus be calculated numerically for any value of the parameters $\sqrt{s}$, $q'_0$, $q'$, $q_0$, $q$, $\hat{q} \cdot \hat{q}'$. There are no complications for $\sqrt{s} < 2m + \mu$, since in that region the function $D$ is negative for all relative energies for which the kernel must be evaluated. On the other hand, above $\sqrt{s} = 2m + \mu$, singularities in the integrand are possible.

The result of the integrations in (3.27) can be partial-wave expanded according to

$$X_\ell(q'_0, q', q_0, q; s) = -\frac{qq'}{8\pi^3} \int_{-1}^{1} dx\, X(q'_\mu, q_\mu; s) P_\ell(x) \,, \tag{3.33}$$

where $x = \hat{q}' \cdot \hat{q}$, using another numerical integration. To obtain a phase shift of the same accuracy as in the ladder case 4 quadratures in each of the Feynman parameter integrations and 3 quadratures in the partial-wave projection are required. There is no need to change any of the quadrature numbers from the ladder solution.

In the case $\mu = m$ our results agree with those of Levine and Wright [38]. In the case of interest to us, $\mu = 0.15$, $\lambda = 0.13$, the crossed-box diagram has a notable contribution. This is illustrated in Fig. 9 where we plot the phase shifts when including both dressing of the propagator and the crossed-box diagram (dotted line). For comparison the phase shifts resulting from the solution of the BSE in the ladder approximation with dressed $\phi$ propagators are also included (dashed line). Comparing the results in Figs. 7 and 9 demonstrates very clearly the relative importance of the crossed-box diagram and the dressing of the $\phi$ propagators. The contribution of the crossed-box diagram is significant in this case despite the small coupling used in the calculation. This suggests that the solution of the $\phi\phi - \sigma\phi\phi$ equations will be substantially different from the ladder result—even below the $\sigma$-production threshold.

## IV. NUMERICAL SOLUTION OF THE $\phi\phi - \sigma\phi\phi$ EQUATIONS

The main motivation for examining the numerical solution of the $\phi\phi - \sigma\phi\phi$ equations is to develop the numerical approximations needed for the solution of the $NN - \pi NN$ equations, where spin and isospin are an additional complication. However, solving the $\phi\phi - \sigma\phi\phi$ equations will also allow a comparison of their solution with the results of the BSE with dressed $\phi$ propagators, and either one-$\sigma$ exchange or one-$\sigma$ exchange plus crossed-box diagram as its kernel. The analysis of Sec. II shows that the solution of the BSE is not going to be identical to that of the coupled-channels problem. However, a comparison may give us some insight into the relative content of the two equations and a measure of how important this difference is. This might tip the balance in favour of one or the other approach when considering $NN$ scattering above the threshold for pion production. Even below the pion-production threshold, recent nucleon-nucleon interactions based on meson exchange have included the crossed pion exchange within the framework of two-body equations rather than in a coupled-channels approach [51,52]. The results of this section might help us determine which of these two methods is best suited for including such crossed diagrams.

In order to simplify matters as far as possible we have assumed that $\phi\sigma$ scattering is dominated by the pole diagram, i.e. $t^{(1)}_{\sigma\phi} = 0$. This yielded the two coupled integral equations (2.10) and (2.11) with the potential $\bar{V}$ of Eq. (2.14). If we also assume that the $\phi - \phi$ interaction is given by one-$\sigma$ exchange and this in turn is represented by an $S$-wave rank-one separable potential, then the input $\phi - \phi$ amplitude $t^{(1)}_{\phi\phi}$ in Eq. (2.10) can be written as in Eq. (3.20). We may therefore recast the coupled equations, Eqs. (2.10) and (2.11), as

$$\tilde{T}_{\phi\phi} = Z_{\phi\phi}\left(1 + \frac{1}{2}d_1 d_2 \tilde{T}_{\phi\phi}\right) + \frac{1}{2} Z_{\phi d}\, \tau_d d_\sigma\, \tilde{T}_{d\phi} \tag{4.1}$$

$$\tilde{T}_{d\phi} = Z_{d\phi} + \frac{1}{2}\, Z_{d\phi}\, d_1 d_2 \tilde{T}_{\phi\phi}\,, \tag{4.2}$$

where $\phi$ refers to the $\phi\phi$ channel, while $d$ refers to the $\sigma(\phi\phi)_d$ channel. Here $\tau_d$ is the propagator for a correlated $\phi\phi$ system. In writing the above coupled-channel equations we have taken into consideration the fact that the $\phi$ is a boson, and the equations need to be symmetrized. Although Eqs. (4.1) and (4.2) can be combined into a single

BSE for $\tilde{T}_{\phi\phi}$, we have chosen to solve the equations as a coupled-channels problem—in anticipation of the fact that if $t^{(1)}_{\sigma\phi} \neq 0$ the coupled equations cannot be reduced to a single BSE.

In what follows all vectors are four-vectors unless otherwise stated. We define the relative four-momentum, $q$, of the $d-\sigma$ state of total four-momentum, $P$, via

$$p_d = \nu_d P + q \; ; \qquad p_\sigma = \nu_\sigma P - q \; , \tag{4.3}$$

where:

$$\nu_d = \frac{s + m_d^2 - \mu^2}{2s} \; ; \qquad \nu_\sigma = \frac{s + \mu^2 - m_d^2}{2s} \; . \tag{4.4}$$

This choice for the relative momentum is made to ensure that when both the $d$ and $\sigma$ particles are on-mass-shell $q_0 = 0$. This in turn guarantees that the pinching of the $q_0''$ integration contour in the BSE for $(\phi\phi)_d\sigma$ scattering will occur at the point $q_0'' = 0$, $q'' = \bar{q}$, where $\bar{q}$ is the magnitude of the on-shell three-momentum. The four-vector $q$ which is defined by (4.3) and (4.4) is known as the Wightman-Gårding relative momentum [53].

With this choice of kinematics we can write the input amplitudes $Z$ as

$$Z_{\phi\phi}(p_1', p_2', p_1, p_2) = -i(2\pi)^4 \delta^{(4)}(p_1' + p_2' - p_1 - p_2) \left[V_\sigma(q - q') + V_\sigma(q + q') - X(q', q; P) - X(q', -q; P)\right] , \tag{4.5}$$

$$Z_{d\phi}(p_1', p_2', p_1, p_2) = -i(2\pi)^4 \delta^{(4)}(p_1' + p_2' - p_1 - p_2) 2\left[V_{d\phi}(q', q; P) + V_{d\phi}(q', -q; P)\right] , \tag{4.6}$$

$$Z_{\phi d}(p_1', p_2', p_1, p_2) = Z_{d\phi}(p_1, p_2, p_1', p_2') \; . \tag{4.7}$$

In Eqs. (4.5) and (4.6), the potentials $V_\sigma$ for $\sigma$-exchange, and $V_{d\phi}$ for $\phi$-exchange, are given by

$$V_\sigma(k) = \frac{g^2}{k^2 - \mu^2} \; , \qquad V_{d\phi}(q', q; P) = g \, \frac{1}{((\nu_d - \frac{1}{2})P + q' - q)^2 - m^2} \, \chi\left(\frac{\nu_\sigma}{2}P + q - \frac{q'}{2}\right) \; . \tag{4.8}$$

The $(\phi\phi)_d$ and $\sigma$ propagators are

$$\tau_d(q; P) = \frac{iS_d((\nu_d P + q)^2)}{(\nu_d P + q)^2 - m_d^2} \; , \qquad d_\sigma(q; P) = \frac{i}{(\nu_\sigma P - q)^2 - \mu^2} \; . \tag{4.9}$$

We now define

$$\tilde{T}_{\phi\phi} = -i(2\pi)^4 \delta^{(4)}(p_1' + p_2' - p_1 - p_2) T_{\phi\phi}(1 + P_{12}) \; , \qquad \tilde{T}_{d\phi} = -i(2\pi)^4 \delta^{(4)}(p_1' + p_2' - p_1 - p_2) T_{d\phi}(1 + P_{12}) \; , \tag{4.10}$$

where $P_{12}$ is the permutation operator for the two $\phi$'s and the factor of $1 + P_{12}$ ensures that scattering takes place in symmetric states only. In particular, no scattering can take place in a state with odd angular momentum. By contrast, if acting on a state with positive parity $1 + P_{12}$ merely gives a factor of two. The equations for these unsymmetrized amplitudes $T_{\phi\phi}$ and $T_{d\phi}$ are then

$$T_{\phi\phi}(q', q; P) = V_{\phi\phi}(q', q; P) + i \int \frac{d^4 q''}{(2\pi)^4} V_{\phi\phi}(q', q''; P) G_\phi(q''; P) T_{\phi\phi}(q'', q; P)$$
$$+ i \int \frac{d^4 q''}{(2\pi)^4} V^S_{\phi d}(q', q''; P) G_d(q''; P) T_{d\phi}(q'', q; P) \tag{4.11}$$

$$T_{d\phi}(q', q; P) = V^S_{d\phi}(q', q; P) + i \int \frac{d^4 q''}{(2\pi)^4} V^S_{d\phi}(q', q''; P) G_\phi(q''; P) T_{\phi\phi}(q'', q; P) \; , \tag{4.12}$$

where $G_\phi$ is defined in Eq. (3.7), and $G_d = -\tau_d d_\sigma$ is the propagator for the $\sigma(\phi\phi)_d$ system. The 'potentials' in Eqs. (4.11) and (4.12) are given by

$$V_{\phi\phi}(q', q; P) = V_\sigma(q' - q) - X(q', q; P) \; , \tag{4.13}$$

$$V^S_{\phi d}(q', q; P) = V_{\phi d}(q', q; P) + V_{\phi d}(-q', q; P) \; , \qquad V^S_{d\phi}(q', q; P) = V_{d\phi}(q', q; P) + V_{d\phi}(q', -q; P) \; . \tag{4.14}$$

If we remove the coupled-channel term $V_{d\phi}$ and the double-counting subtraction $X$ the first equation here reduces to the Bethe-Salpeter equation solved for $\phi\phi$ scattering in Sec. III A, thus justifying our use of the distinguishable-particle BSE to determine the $\phi\phi \to \phi\phi$ amplitude in the allowed two-body channels.

Having derived these coupled integral equations they must be recast into a form convenient for computation. This requires three steps. Firstly, a partial-wave expansion must be performed in order to reduce the dimensionality of the equations from four to two. Secondly, any pinches causing difficulties in the energy range of interest must be removed. (Note that such pinches are *not* removed by Wick rotation.) Once this is done Wick rotation can be performed, thus moving the region of integration away from the poles of the $\phi$ propagators and exchange diagrams.

## A. Partial-wave expansion

The partial-wave expansion is performed exactly as defined in Sec. III A. The amplitudes $T_{\phi\phi}$ and $T_{d\phi}$ and the "potentials" $V_\sigma$, $X$, $V_{d\phi}$ and $V_{\phi d}$ are expanded according to Eq. (3.5). This produces the results given in Eq. (3.8) and (3.33) for $V_\sigma$ and $X$. For $V_{d\phi}$ partial fractions may be used to split the integration, leading to:

$$V_{d\phi;\ell}(q'_0, q', q_0, q; s) = \frac{\sqrt{\lambda}}{\pi^2} \frac{1}{D} \left(Q_\ell(\gamma_{d\phi}) - Q_\ell(\gamma_\beta)\right) , \tag{4.15}$$

$$D = 2q'q(\gamma_{d\phi} - \gamma_\beta) , \tag{4.16}$$

$$\gamma_{d\phi} = \frac{q^2 + q'^2 + m^2 - (q'_0 - q_0 + (\nu_d - \frac{1}{2})\sqrt{s})^2}{2q'q} , \qquad \gamma_\beta = \frac{\frac{q'^2}{4} + q^2 + \beta^2 - (\frac{\nu_\sigma\sqrt{s}}{2} + q_0 - \frac{q'_0}{2})^2}{q'q} . \tag{4.17}$$

With this result in hand it is easy to compute $V_{\phi d;\ell}$, since

$$V_{\phi d;\ell}(q'_0, q', q_0, q; s) = V_{d\phi;\ell}(q_0, q, q'_0, q'; s) . \tag{4.18}$$

The use of these partial-wave expansions leads to a set of coupled two-dimensional integral equations for the half-off-shell partial-wave amplitudes $T_{\phi\phi;\ell}$ and $T_{d\phi;\ell}$. These equations must then have all pinches removed.

## B. Kowalski-Noyes reduction

Although the main interest in a coupled-channels approach is to investigate scattering in the energy region where three-body unitarity plays a central role, at this stage we are going to restrict our analysis to the energy region below the three-body unitarity cut. This is partly because in this energy domain we can use the Kowalski-Noyes approach, discussed in the last section, to remove the two-body unitarity cut.

Since the coupled-channel equations do not couple amplitudes of different angular momentum, we now simplify the notation by dropping the angular momentum label on all amplitudes.

The Kowalski-Noyes reduction is carried out for the two amplitudes in question by making the factorizations

$$T_{\phi\phi}(q'_0, q', 0, \bar{q}; s) = f_{\phi\phi}(q'_0, q') T^{on}_{\phi\phi}(s) , \qquad T_{d\phi}(q'_0, q', 0, \bar{q}; s) = f_{d\phi}(q'_0, q') T^{on}_{\phi\phi}(s) , \tag{4.19}$$

where the dependence of the Kowalski-Noyes half-off-shell functions on $s$ has been suppressed.

Using these factorizations the partial-wave-expanded coupled-channel equations for the KN half-off-shell functions become

$$f_{\phi\phi}(q'_0, q') = \frac{V_{\phi\phi}(q'_0, q', 0, \bar{q}; s)}{V^{on}_{\phi\phi}} - i \int_{-\infty}^{\infty} dq''_0 \int_0^\infty dq'' \, \tilde{V}_{\phi\phi}(q'_0, q', q''_0, q''; s) \, G_\phi(q''_0, q''; s) \, f_{\phi\phi}(q''_0, q'')$$

$$- i \int_{-\infty}^{\infty} dq''_0 \int_0^\infty dq'' \, \tilde{V}_{\phi d}(q'_0, q', q''_0, q''; s) \, G_d(q''_0, q''; s) \, f_{d\phi}(q''_0, q'') , \tag{4.20}$$

$$f_{d\phi}(q'_0, q') = \frac{V^S_{d\phi}(q'_0, q', 0, \bar{q}; s)}{V^{on}_{\phi\phi}} - i \int_{-\infty}^{\infty} dq''_0 \int_0^\infty dq'' \, \tilde{V}_{d\phi}(q'_0, q', q''_0, q''; s) \, G_\phi(q''_0, q''; s) \, f_{\phi\phi}(q''_0, q'')$$

$$- i \int_{-\infty}^{\infty} dq''_0 \int_0^\infty dq'' \, \tilde{V}_{dd}(q'_0, q', q''_0, q''; s) \, G_d(q''_0, q''; s) \, f_{d\phi}(q''_0, q'') , \tag{4.21}$$

where:

$$\tilde{V}_{\phi\phi}(q'_0, q', q_0, q; s) = V_{\phi\phi}(q'_0, q', q_0, q; s) - \frac{V_{\phi\phi}(q'_0, q', 0, \bar{q}; s) \, V_{\phi\phi}(0, \bar{q}, q_0, q; s)}{V^{on}_{\phi\phi}} , \tag{4.22}$$

$$\tilde{V}_{\phi d}(q'_0, q, q_0, q; s) = V^S_{\phi d}(q'_0, q', q_0, q; s) - \frac{V_{\phi\phi}(q'_0, q', 0, \bar{q}; s) \, V^S_{\phi d}(0, \bar{q}, q_0, q; s)}{V^{on}_{\phi\phi}} , \tag{4.23}$$

$$\tilde{V}_{d\phi}(q_0', q, q_0, q; s) = V_{d\phi}^S(q_0', q', q_0, q; s) - \frac{V_{d\phi}^S(q_0', q', 0, \bar{q}; s)\, V_{\phi\phi}(0, \bar{q}, q_0, q; s)}{V_{\phi\phi}^{on}} \;, \qquad (4.24)$$

$$\tilde{V}_{dd}^L(q_0', q, q_0, q; s) = -\frac{V_{d\phi}^S(q_0', q', 0, \bar{q}; s) V_{\phi d}^S(0, \bar{q}, q_0, q; s)}{V_{\phi\phi}^{on}} \;. \qquad (4.25)$$

It is clear from the structure of these $\tilde{V}$s that the pinch in $G_\phi(q_0'', q''; s)$ is now always ameliorated by a zero in $\tilde{V}$.

In order to obtain phase shifts it is necessary to calculate the on-shell $t$-matrix, $T_{\phi\phi}(s)$, via

$$T_{\phi\phi}^{on}(s) = \frac{V_{\phi\phi}^{on}(s)}{1 + iI_\phi(s) + iI_d(s)} \;, \qquad (4.26)$$

with:

$$I_\phi(s) = \int_{-\infty}^{\infty} dq_0'' \int_0^{\infty} dq''\, V_{\phi\phi}(0, \bar{q}, q_0'', q''; s)\, G_\phi(q_0'', q''; s)\, f_{\phi\phi}(q_0'', q'') \;, \qquad (4.27)$$

$$I_d(s) = \int_{-\infty}^{\infty} dq_0'' \int_0^{\infty} dq''\, V_{\phi d}(0, \bar{q}, q_0'', q''; s)\, G_d(q_0'', q''; s)\, f_{d\phi}(q_0'', q'') \;. \qquad (4.28)$$

### C. Wick rotation

By writing coupled integral equations for the KN functions we have overcome the problem of the pinching between the poles of the $\phi$ propagator. However, to convert the equations to matrix equations we have to avoid the other singularities of the kernel. This can be achieved by performing a Wick rotation in the $q_0''$ integrals. Such a rotation requires that we take into consideration the singularities of both the kernel and the amplitude in the first and third quadrant of the $q_0''$-plane. These singularities have been examined in detail [28] with the result that for $2m < \sqrt{s} < m_d + \mu$ the only analytic structure which may cause problems during the rotation are the two poles from the Green's function $G_\phi$.

The final equations after Wick rotation can thus be written as

$$f_{\phi\phi}(iq_0', q') = \frac{V_{\phi\phi}(iq_0', q', 0, \bar{q}; s)}{V_{\phi\phi}^{on}} + \int_0^{\infty} dq_0'' \int_0^{\infty} dq''\, \tilde{V}_{\phi\phi}^S(iq_0', q', iq_0'', q'')\, G_\phi(iq_0'', q''; s)\, f_{\phi\phi}(iq_0'', q'')$$

$$- \pi \int_0^{\bar{q}} dq''\, \tilde{V}_{\phi\phi}^S(iq_0', q', \bar{w}(q''), q'')\, \frac{1}{2\sqrt{s}\bar{w}(q'')E_\phi(q'')}\, g_{\phi\phi}(q'')$$

$$+ \int_{-\infty}^{\infty} dq_0'' \int_0^{\infty} dq''\, \tilde{V}_{\phi d}(iq_0', q', iq_0'', q''; s)\, G_d(iq_0'', q''; s)\, f_{d\phi}(iq_0'', q'') \;, \qquad (4.29)$$

$$f_{d\phi}(iq_0', q') = \frac{V_{d\phi}^S(iq_0', q', 0, \bar{q}; s)}{V_{\phi\phi}^{on}} + \int_0^{\infty} dq_0'' \int_0^{\infty} dq''\, \tilde{V}_{d\phi}^S(iq_0', q', iq_0'', q''; s)\, G_\phi(iq_0'', q''; s)\, f_{\phi\phi}(iq_0'', q'')$$

$$- \pi \int_0^{\bar{q}} dq''\, \tilde{V}_{d\phi}^S(iq_0', q', \bar{w}(q''), q'')\, \frac{1}{2\sqrt{s}\bar{w}(q'')E_\phi(q'')}\, g_{\phi\phi}(q'')$$

$$+ \int_{-\infty}^{\infty} dq_0'' \int_0^{\infty} dq''\, \tilde{V}_{dd}(iq_0', q', iq_0'', q''; s)\, G_d(iq_0'', q''; s)\, f_{d\phi}(iq_0'', q'') \;, \qquad (4.30)$$

where in order to simplify the equations we have assumed undressed particles in taking the residue of $G_\phi$, but the appropriate modification for dressed particles may easily be made. Also:

$$\tilde{V}^S_{\phi\phi}(q'_0, q', q''_0, q''; s) = \tilde{V}_{\phi\phi}(q'_0, q', q''_0, q''; s) + \tilde{V}_{\phi\phi}(q'_0, q', -q''_0, q''; s) , \qquad (4.31)$$

$$\tilde{V}^S_{d\phi}(q'_0, q', q''_0, q''; s) = \tilde{V}_{d\phi}(q'_0, q', q''_0, q''; s) + \tilde{V}_{d\phi}(q'_0, q', -q''_0, q''; s) = 2\tilde{V}_{d\phi}(q'_0, q', q''_0, q''; s) . \qquad (4.32)$$

Note that $g_{\phi\phi}(q') \equiv f_{\phi\phi}(\bar{w}(q'), q')$ obeys the equation

$$g_{\phi\phi}(q') = \frac{V_{\phi\phi}(\bar{w}(q'), q', 0, \bar{q}; s)}{V^{on}_{\phi\phi}} + \int_0^\infty dq''_0 \int_0^\infty dq'' \, \tilde{V}^S_{\phi\phi}(\bar{w}(q'), q', iq''_0, q''; s) \, G_\phi(iq''_0, q''; s) \, f_{\phi\phi}(iq''_0, q'')$$

$$-\pi \int_0^{\bar{q}} dq'' \, \tilde{V}^S_{\phi\phi}(\bar{w}(q'), q', \bar{w}(q''), q''; s) \, \frac{1}{2\sqrt{s}\bar{w}(q'')E_\phi(q'')} \, g_{\phi\phi}(q'')$$

$$+ \int_{-\infty}^\infty dq''_0 \int_0^\infty dq'' \, \tilde{V}_{\phi d}(\bar{w}(q'), q', iq''_0, q''; s) \, G_d(iq''_0, q''; s) \, f_{d\phi}(iq''_0, q'') . \qquad (4.33)$$

Wick rotation is also necessary to remove the poles in the integrals in Eq. (4.26). Above we saw that

$$I_\phi(s) = 2a(s) + b(s) , \qquad (4.34)$$

where $a(s)$ and $b(s)$ are given by Eqs. (3.15) and (3.16), but with $V_{\phi\phi}$ and $f_{\phi\phi}$ replacing $V$ and $f$. As for $I_d$, by Wick rotating the $q''_0$ integration in Eq. (4.28) the result

$$iI_d(s) = -\int_{-\infty}^\infty dq''_0 \int_0^\infty dq'' \, V_{\phi d}(0, \bar{q}, iq''_0, q'') \, G_d(iq''_0, q''; s) f_{d\phi}(iq''_0, q''; s) \qquad (4.35)$$

is found. In contrast to the BSE equation, where we had two coupled integral equations for the KN functions, here we have three coupled integral equations, with the $\sigma(\phi\phi)_d$ channel being the source of the extra equation.

### D. Numerical results

The equations (4.29), (4.30) and (4.33) are in a form suitable for computation, having had almost all poles and cuts removed from the kernel. The only remaining pole is that at $q''_0 = 0$, $q'' = \bar{q}$, due to the zero in $\tilde{V}$ at that point being only first order, while the pole in $G$ at that point is second order. However, as in the ladder BSE case, this pole may be dealt with by making a change of integration variable and choosing appropriate quadratures.

The equations (4.29), (4.30) and (4.33) are solved for the Kowalski-Noyes half-off-shell functions and the results used to calculate phase shifts according to Eqs. (3.10), (4.26), (3.15), (3.16) and (4.35). The equations are solved by discretizing the integrals using Gauss-Legendre quadratures and then applying matrix inversion. In order to guarantee phase shift accuracy to three significant figures 14 and 22 quadratures are necessary for the $q''_0$ and $q''$ $\phi\phi$ intermediate-state integration, while 44 and 14 quadratures are necessary in the $q''_0$ and $q''$ integrations over the $(\phi\phi)_d\sigma$ intermediate-state. (Note that the $q''_0$ integration is from $-\infty$ to $\infty$ in the $(\phi\phi)_d\sigma$ case.) 8 quadratures are needed for the $q''$ integration over $[0, \bar{q}]$.

As a first step in the application of the coupled-channel equations, we examine the contribution of the double-counting subtraction. In Fig. 10 we present the phase shifts for $\phi - \phi$ scattering with (solid line labeled $CC - X$) and without (dash-dot line labeled $CC$) this subtraction. Included for comparison are the phase shifts resulting from the solution of the BSE with (dotted line labeled Ladder+$X$) and without (dashed line labeled Ladder) the crossed-box diagram. All four curves are calculated with the $\phi$ propagators dressed as described in Section III C.

Firstly, observe that the double-counting subtraction has a notable but not substantial effect on the phase shifts. This is a measure of the contribution of the crossed-box diagram, and the results are consistent with the difference between the ladder and ladder plus crossed-box diagram BSE calculations. Secondly, it is easily seen from Fig. 10 that there is a considerable gap between the phase shifts given by the ladder plus crossed-box calculation (Ladder + $X$) and those produced by the coupled-channels calculation with the double-counting subtraction ($CC - X$). The difference between these two calculations is that $CC - X$ contains the complete coupling to the $(\phi\phi)_d\sigma$ channel in its kernel, whereas Ladder + $X$ includes only two Feynman graphs in its kernel. Thus we would expect some gap between the phase shifts predicted by these two calculations. However, at first sight it is odd that such a large difference could occur below the $\sigma$-production threshold, and with the small coupling strength, $\lambda = 0.13$.

### E. Effects of vertex dressing

The explanation of the large difference could be due to the two diagrams $(a)$ and $(b)$ of Fig. 2. These graphs are effectively included in the coupled-channels calculation, thus partially dressing the $\phi\phi\sigma$ vertices. By contrast, the vertices used in the ladder and ladder plus crossed-box calculations were undressed.

To gauge the size of this effect, we introduce vertex dressing in the ladder BSE and adjust the coupling strength $\lambda$ to get a binding for the $\phi - \phi$ state that is comparable with the deuteron binding energy. We then use a separable approximation to this new ladder BSE amplitude as input to the coupled-channel equations. In this way diagrams $(a)$ and $(b)$ of Fig. 2 are included in both the BSE and coupled-channel equations. The effect of this on the coupled-channel equations is that some vertices in Fig. 2 are now bare vertices, and some are dressed vertices.

We begin by seeking a self-consistent equation for the one-$\sigma$-exchange piece of the $CC-X$ calculation. In other words, we must determine what form the $\sigma\phi\phi$ vertex function must have if it is to be the vertex in both the input $\phi\phi$ ladder interaction *and* the one-$\sigma$-exchange piece of the $\phi\phi$ interaction which results from the coupled-channels approach. It is easily demonstrated that for this to be the case the vertex function $f$ must obey

$$f(p'_\nu, p_\nu, k_\nu) = g_b + ig_b^2 \int \frac{d^4k'}{(2\pi)^4} \frac{1}{k'_\nu k'^\nu - \mu^2} \frac{1}{(p'-k')_\nu(p'-k')^\nu - m^2} f(p'_\nu - k'_\nu, p_\nu - k'_\nu, k_\nu) \frac{1}{(p-k')_\nu(p-k')^\nu - m^2}. \tag{4.36}$$

(This equation is given in diagrammatic form in Fig. 11.) In general, equation (4.36) is difficult to solve. On the other hand, a reasonable approximation is obtained by arguing that the $f$ in the integrand may be approximated by $g$, the dressed coupling constant. Such an approximation includes the one-loop dressing of the vertex exactly, and higher-loop dressings approximately. Its use may be justified by appeal to the same argument which validated our use of one-loop propagator dressing above. Higher-loop dressings make contributions to $\phi\phi\sigma\sigma$ and higher-state unitarity. The one-loop result is therefore all we need to enforce $\phi\phi\sigma$ unitarity. Moreover, due to the small value of the coupling being used we might expect that any error in such an approximation is small.

The integration over four-momentum $k'_\nu$ may now be done via the Feynman technique, and the result is

$$f(p'_\nu, p_\nu, k_\nu) = g_b - g\lambda_b \int_0^1 d\alpha \int_0^{1-\alpha} d\beta \frac{1}{D(\alpha, \beta | p'_\mu, p_\mu, k_\mu)}, \tag{4.37}$$

$$D(\alpha, \beta | p'_\mu, p_\mu, k_\mu) = \beta(1 - \alpha - \beta)p'_\mu p'^\mu + \alpha\beta k_\mu k^\mu$$
$$+ \alpha(1 - \alpha - \beta)p_\mu p^\mu - (m^2 - \mu^2)(\alpha + \beta) - \mu^2, \tag{4.38}$$

where $p_\mu$ is the initial nucleon four-momentum, $p'_\mu$ the final nucleon four-momentum, $k_\mu$ the pion four-momentum, and $\lambda_b = \frac{g_b^2}{16\pi^2}$, with $g_b$ the bare coupling constant.

Defining $I$ to be the value of the integral for $p^2 = p'^2 = m^2$ and $k^2 = \mu^2$ (without the factors of $g$ and $g_b$ included), suggests that the dressed and bare coupling constants are related by

$$g = \frac{g_b}{1 + g_b^2 \frac{I}{16\pi^2}}. \tag{4.39}$$

When the integral $I$ is evaluated numerically we find $I = -1.719$.

To include such dressing for the vertices in any of our previous BSE calculations only requires a change in the definition of the "potential" $V$. The additional vertex dressing does not interfere with the Wick rotation, since none of its cuts intrude into the first and third quadrants for $\sqrt{s} < 2m + \mu$. The momentum-dependence of the vertices may be calculated straightforwardly up to $\sqrt{s} = 2m + \mu$, since at these energies the integrand in Eq. (4.37) is regular for all four-momenta used in the kernel of the integral equations for the KN functions.

Numerical partial-wave expansion is used to evaluate the one-$\sigma$-exchange interaction with dressed vertices. To obtain an accuracy of 3 significant figures in the phase shifts and 4 significant figures in the squares of the bound-state masses, 8 quadratures in this partial-wave integration and 4 quadratures in the Feynman integral is sufficient. The dressing of the $\phi$ propagator is done as before but with a strength $\sqrt{\lambda\lambda_b}$, rather than $\lambda$, in order to simulate the effect of dressing one of the two vertices in the dressing loop.

Firstly we consider the bound-state calculation. Tests show that the discretization of the integral equation can be carried out with the same number of quadratures without the accuracy diminishing. It is found that the bare coupling required for the scalar "deuteron" to have the desired mass is $\lambda_b = 0.0935$, corresponding to $\lambda = 0.133$.

To examine the importance of the momentum dependence of the vertex dressing, in Fig. 12 we compare the phase shifts with dressed propagators and vertices (solid line) with the a ladder calculation in which there is no vertex dressing and strength of the bare coupling is taken to be $\lambda_b = \lambda = 0.133$ (dashed line). These results show clearly that the momentum dependence introduced as a result of the dressing only has a small effect on the final phase shifts. In fact, it reduces the phase shifts and to that extent increases the difference between these results and the results of the coupled-channels calculation reported in Fig. 10. The small effect of the momentum dependence on the phase shifts provides justification for the approximation used in solving (4.37).

We can now use the results of the BSE with dressed vertices and propagators as input to the coupled-channel equations. To do this, we need to fit the solution of the BSE in the ladder approximation with dressed vertices and propagators with a separable potential. The result of this fit is presented in Fig. 13. The best fit is achieved with $\beta = 0.46760$, which is not very different from the $\beta$ found in the undressed-vertex case.

This separable expansion of the ladder BSE with dressed vertices can now be used as the input to the coupled-channels calculation. However, if this is done additional double-counting is introduced, since the Feynman graph in Fig. 14 is double-counted, and so a further subtraction in $V_{\phi\phi}$ is required.

Other complications arise because the vertices which produce the coupling to the $\sigma(\phi\phi)_d$ channel should be dressed. However, there is no $\phi\sigma$ non-pole t-matrix, and therefore no way for these vertices to be dressed in the coupled-channels theory derived above. To cover this deficiency of our calculation we arbitrarily replace $g_b$ by $g$ everywhere in $V_{\phi d}$ and $V_{d\phi}$.

The problem with this *ad hoc* solution is that the $(\phi\phi)_d\sigma$ coupling vertices which lead to vertex dressing for the one-$\sigma$-exchange $\phi\phi$ interaction are now also dressed. Therefore corrections are made to the interaction $V_{\phi\phi}$, in order to remove all $g$'s which lead to vertex dressing and replace them with $g_b$'s once again.

Note: dressed vertices in the crossed-box diagram, the $\phi$ propagator dressing loop, and the "true" coupling to the $(\phi\phi)_d\sigma$ channel all have *no* momentum-dependence—they merely contain the coupling constant $g$ rather than $g_b$. We think this is a reasonable approximation to the full result, since the ladder results show that, for the values of the coupling of interest here, it is the overall strength of the coupling which is the major effect, not its momentum-dependence.

Once these corrections to $V_{\phi\phi}$, $V_{\phi d}$, and $V_{d\phi}$ are made the calculation proceeds exactly as above. The number of quadratures required for three significant figure accuracy remains unchanged. The results of these calculations are shown in Fig. 15. Coupled-channels calculations both with and without the crossed-box diagram subtraction are compared to the ladder calculation with dressed vertices and the same calculation with the crossed-box diagram added. Now that graphs like (*a*) and (*b*) of Fig. 2, but with bare vertices, are included in the kernel of the BSE, the $CC - X$ and Ladder + $X$ calculations contain the same second and fourth-order Feynman graphs. Therefore, it is the two middle curves of Fig. 15 which should be compared. These two curves are much closer together than was the case in Fig. 10, thus indicating that the discrepant vertex dressing was the main reason for the large gap between the same curves in that plot.

So, we conclude that if the vertex dressing is done approximately consistently in the Ladder + $X$ and $CC - X$ calculations then the resultant phase shift curves lie very close together. The double-counting subtraction is crucial to our obtaining this agreement.

## V. CONCLUSION

In this paper a coupled-channels formulation of $\phi\phi$ scattering in a $\phi^2\sigma$ field theory was investigated. In particular, we found that the $\phi\phi - \sigma\phi\phi$ equations can be solved using an extension of the method developed by Levine *et al* [45] for the ladder Bethe-Salpeter equation (BSE). This may be done in a moderate amount of computer time on a SUN workstation. Since these equations are just the $NN - \pi NN$ equations of Ref. [27] adapted to a scalar field theory with $t^{(1)}_{\phi\sigma} = 0$ this indicates that it is feasible to solve the four-dimensional covariant $NN - \pi NN$ equations numerically. It also yields phase shifts for $\phi\phi$ scattering up to the $(\phi\phi)_d\sigma$ threshold, thereby showing that the $\phi\phi - \sigma\phi\phi$ equations can be used to effectively include in the BSE kernel the infinitely many diagrams involving one explicit $\sigma$. So, it allows a comparison between such a coupled-channels ($CC$) approach to $\phi\phi$ scattering and descriptions using a single BSE, which allow only a finite number of Feynman graphs to be included in the kernel.

As discussed at length in Refs. [8,27,28], in most previous four-dimensional $NN - \pi NN$ equations some diagrams are included more than once in the kernel of the coupled-channel equations. In the first calculation of Sec. IV D the only double-counting removal required is that of the crossed-box diagram. Its subtraction is found to make a significant difference to the phase shifts obtained from the coupled-channels calculation.

However, the gap between the phase shifts produced by this, $CC - X$, calculation and those obtained when the crossed-box diagram is added to the kernel of the ladder BSE (Ladder + $X$) is surprisingly large. Closer examination reveals that the $\phi\phi\sigma$ vertices in the two approaches are not dressed in the same way. The coupling to the $(\phi\phi)_d\sigma$ channel in the $CC$ formulation introduces some dressing of the vertices. If a direct comparison is to be made this dressing must be included in the ordinary BSE calculations. Once this consistent dressing is implemented a striking decrease in the gap between the $CC-X$ and Ladder $+ X$ calculations occurs.

It is then seen that the "true" coupling to the $(\phi\phi)_d\sigma$ channel makes little difference to the phase shifts. In other words, if all of the second and fourth order diagrams which are effectively included in the $CC-X$ calculation's kernel are summed in the kernel of the single BSE then the results of the two calculations are very similar. It is worth noting that the agreement between the single and coupled BSE approaches is destroyed by the omission of any fourth-order diagram from the Ladder $+ X$ calculation, and by the over-counting of the fourth-order crossed-box graph in the $CC$ calculation. Whether this close agreement persists above the $(\phi\phi)_d\sigma$ threshold remains to be investigated.

It is only because the "true" $(\phi\phi)_d\sigma$ channel is relatively unimportant that we can claim to have summed the field theory correctly. If this were not the case our use of the ladder BSE as the basis for the input $\phi\phi$ interaction would be open to criticism. Strictly speaking in order to "bootstrap" the theory up we should construct a parameterization of the $\phi\phi$ amplitude resulting from our $CC-X$ calculation and use this as input to a *new* $CC-X$ calculation, repeating this process until convergence is obtained. However, if this procedure were implemented additional double-counting would be introduced into the equations. This would have to be explicitly removed. As the "true" $(\phi\phi)_d\sigma$ coupling appears fairly small in our calculation, we do not pursue such a bootstrapping procedure here.

In this paper we have only calculated $\phi\phi$ scattering up to the $\sigma$-"deuteron" threshold. In modifying the methods of this paper for work above $\sqrt{s} = m_d + \mu$ three issues arise. Firstly, once this energy is reached the $\sigma$ and $d$ poles pinch each other, thus generating the elastic $(\phi\phi)_d\sigma$ threshold. This pinch and threshold may be removed by modifying the Kowalski-Noyes method so that it applies to coupled-channels problems. Secondly, the validity and efficacy of Wick rotation must be examined. The discussion of analytic structure in Appendix G of Ref. [28] shows that, for $m_d + \mu \leq \sqrt{s} < 2m + 2\mu$, the only cut which threatens Wick rotation is the one which represents the process $(\phi\phi)_d \to \phi+\phi$. However, even if $\sqrt{s}$ is large enough for this "deuteron" break-up cut to intrude into the third quadrant the cut may still be avoided by Wick rotating about a point on the negative $q_0''$ axis in $(\phi\phi)_d\sigma$ intermediate states. Provided this point is suitably chosen, the only structure which can obstruct such a rotation is the $\sigma$-propagator pole. Hence an auxiliary equation for the Kowalski-Noyes functions at $q_0'' = \nu_\sigma\sqrt{s} - E_\sigma(q'')$ must be written when the Wick rotation is performed. Once the rotation is completed the only new analytic structure in the kernel occurs in that part of the kernel corresponding to the transition $(\phi\phi)_d + \sigma \to \phi + \phi$. In the auxiliary equation for $f_{\phi\phi}$ at $\bar{\omega}(q')$ that piece of the kernel acquires logarithmic branch-points, due to the possibility of real $\sigma$ production via the diagram shown in Fig. 16. Thus, Wick rotation is permitted and succeeds in eliminating most of the troublesome analytic structure from the kernel. Finally, the increase in available energy leads to complications in the calculation of Feynman integrals. In both the crossed-box graph and the vertex loop some of the intermediate-state particles can propagate on-shell once $\sqrt{s}$ reaches $2m + \mu$. So, singularities occur in the integrands of the Feynman integrals and the integrals can no longer be done in a straightforward way.

Another point for future investigation is the inclusion of a non-pole $\phi\sigma$ interaction in the theory. This might be done by solving the ladder BSE for unequal mass particles using the driving term shown in Fig. 17. The resultant amplitude could then be parameterized by a separable interaction and used as input to the $\phi\phi - \sigma\phi\phi$ equations. This obviously increases the amount of physics summed in the theory, but also has three other effects. Firstly, by introducing an additional channel, $(\phi\sigma) + \phi$, the computer time required is raised. Secondly, the vertex dressing problem no longer needs to be remedied "by hand" since all vertex dressing will be done in a consistent way. Finally, further double-counting corrections need to be introduced, as discussed in Section II. However, since all of these further double-counting corrections involve the removal of the crossed diagram from the $\phi\sigma$ interaction this third point should not complicate matters greatly.

Few changes will be needed in order to apply the methods of this paper to the $NN - \pi NN$ system. The first step in such a calculation will be to perform the spin and isospin algebra involved in the partial-wave decomposition of the four-dimensional $NN - \pi NN$ equations. Once this is done a set of coupled partial-wave expanded equations of the form solved in this paper is obtained. Questions remain about the validity of Wick rotation in the $NN - \pi NN$ case, since some of the propagators now are proportional to $\frac{1}{q}$, rather than the $\frac{1}{q^2}$ of the scalar case. But, the presence of form factors in our equations will ensure that the integrands go to zero fast enough to validate the use of Wick rotation. Thus, once a set of coupled equations for the $NN - \pi NN$ system has been obtained the approach described here may be applied. The computing time required for such a calculation will obviously be much longer than for those in the scalar field theory. It should also be pointed out that if results for the $NN - \pi NN$ system above the two-pion threshold are needed a method which is more sophisticated than our "naive" Wick rotation should be used.

However, for calculations up to the second production threshold the work of this paper shows that the numerical solution of the coupled field-theoretic equations derived in Ref. [27] is entirely possible.

## ACKNOWLEDGMENTS

We thank C. H. M. van Antwerpen, who wrote the code which was the basis for our ladder BSE code. We are also grateful to B. C. Pearce for suggesting the alternative subtraction technique in the calculation of $a$ and $b$ in Eq. (3.14). This work was conducted while D. R. P. held an Australian Postgraduate Award. We acknowledge the financial support of the Australian Research Council and the United States Department of Energy (Contract no. DE-FG02-93ER-40762).

FIG. 1. The subtractions in $\bar{V}$ required to avoid double counting.

FIG. 2. The lowest-order Feynman diagrams that contribute to $V_{eff}$ if the input $\phi-\phi$ amplitude is the solution to the ladder BSE.

FIG. 3. A pictorial representation of the ladder Bethe-Salpeter equation as given in Eq. (3.2).

FIG. 4. Coupling constant versus bound-state position for the ladder BSE with $m=1$ and $\mu=0.15$.

FIG. 5. Comparison of the separable approximation (dashed line) to the ladder calculation with undressed particles (solid line) for $\mu = 0.15$ and $\lambda = 0.13$.

FIG. 6. The Schwinger-Dyson equation for the $\phi$-particle propagator. Heavy lines represent dressed propagators, and the lighter lines represent bare propagators.

FIG. 7. Real part of the $S$-wave phase shifts in the ladder approximation, with (solid curve) and without (dashed curve) one-loop dressing for the case $\lambda = 0.13$.

FIG. 8. Imaginary part of the $S$-wave phase shifts in the ladder approximation, with (solid curve) and without (dashed curve) one-loop dressing for the case $\lambda = 0.13$.

FIG. 9. Real part of the $S$-wave phase shifts with (dotted curve) and without (dashed curve) the crossed-box diagram, for the case of dressed particles with $\lambda = 0.13$.

FIG. 10. Comparison of four different ways of calculating the real part of the $S$-wave phase shifts. The dashed curve is the result of the ladder calculation, the dotted curve includes the crossed-box diagram, the dot-dashed curve is the straightforward coupled-channels result, and the solid curve is the coupled-channels result with the crossed-box diagram subtracted so as to remove double-counting.

FIG. 11. The non-linear equation for the vertex function $f$.

FIG. 12. The real part of the $S$-wave phase shifts for the case $\lambda_b = 0.0935$ with vertex dressing included (solid line), compared with the case $\lambda = 0.133$ with no vertex dressing (dashed line).

FIG. 13. The real part of the $S$-wave phase shifts for the case $\lambda_b = 0.0935$ with vertex dressing included (solid line), compared with a separable approximation to this result (dashed line).

FIG. 14. A Feynman graph which is double-counted in the coupled-channels calculation if an input $\phi - \phi$ amplitude with dressed vertices is used.

FIG. 15. Comparison of single-BSE approach, with vertices approximately dressed to all loop orders, against coupled-channels approach. Legend as in Fig. 10.

FIG. 16. The diagram which, above $\sigma$-production threshold, leads to logarithmic branch-points in the kernel of the Wick-rotated coupled integral equations.

FIG. 17. A possible driving term for the ladder BSE for $\phi\sigma$ scattering.

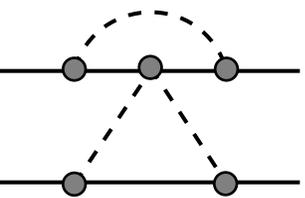 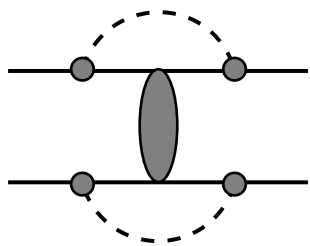 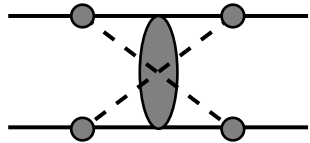 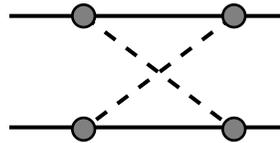

$D_2$      $D_\sigma$      $X$      $B$

$\overline{V}$ +

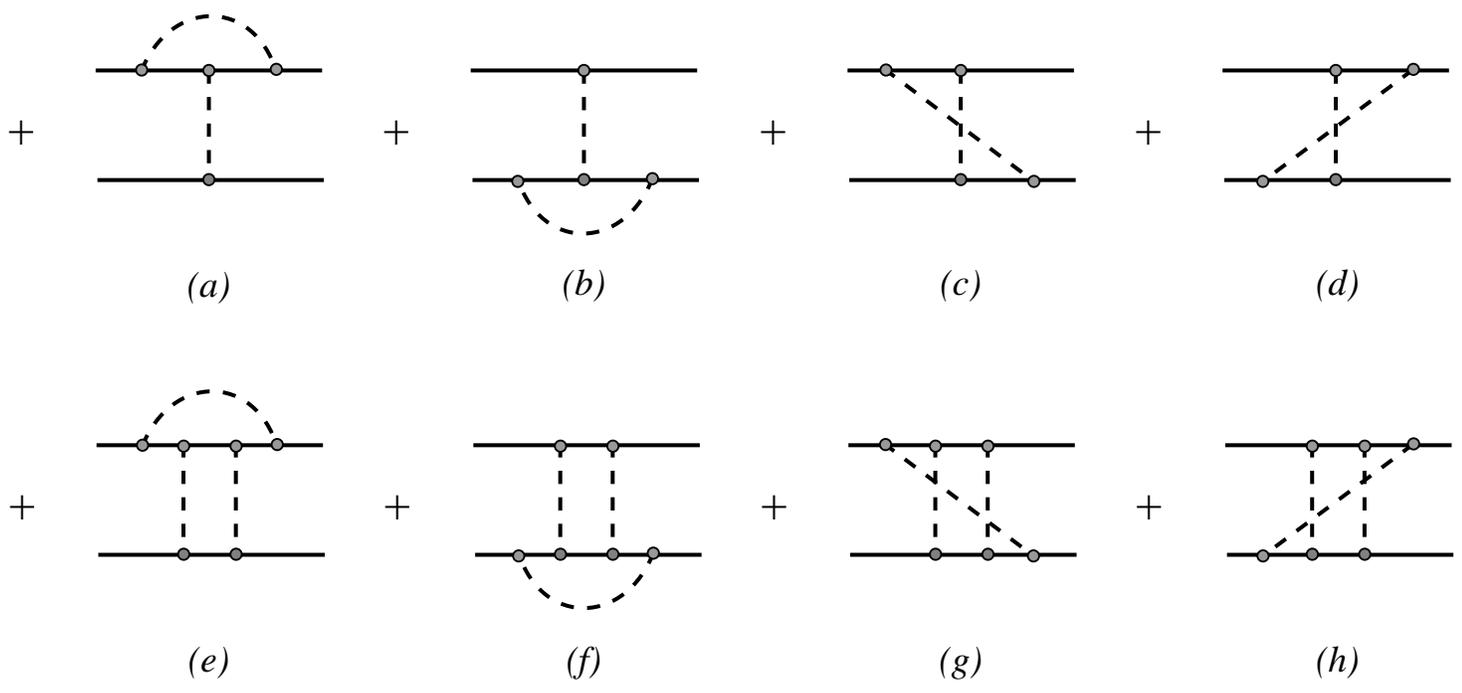

(a)     (b)     (c)     (d)

(e)     (f)     (g)     (h)

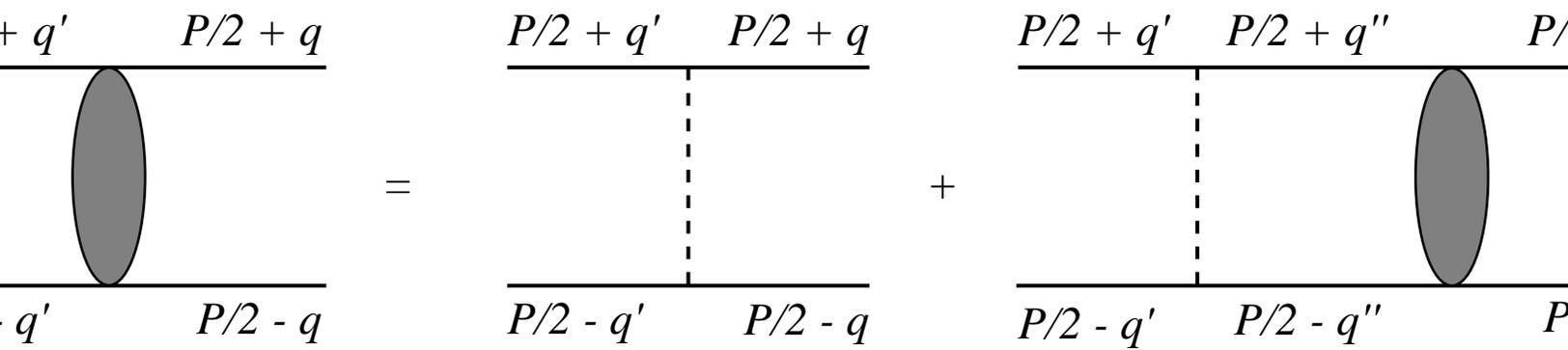

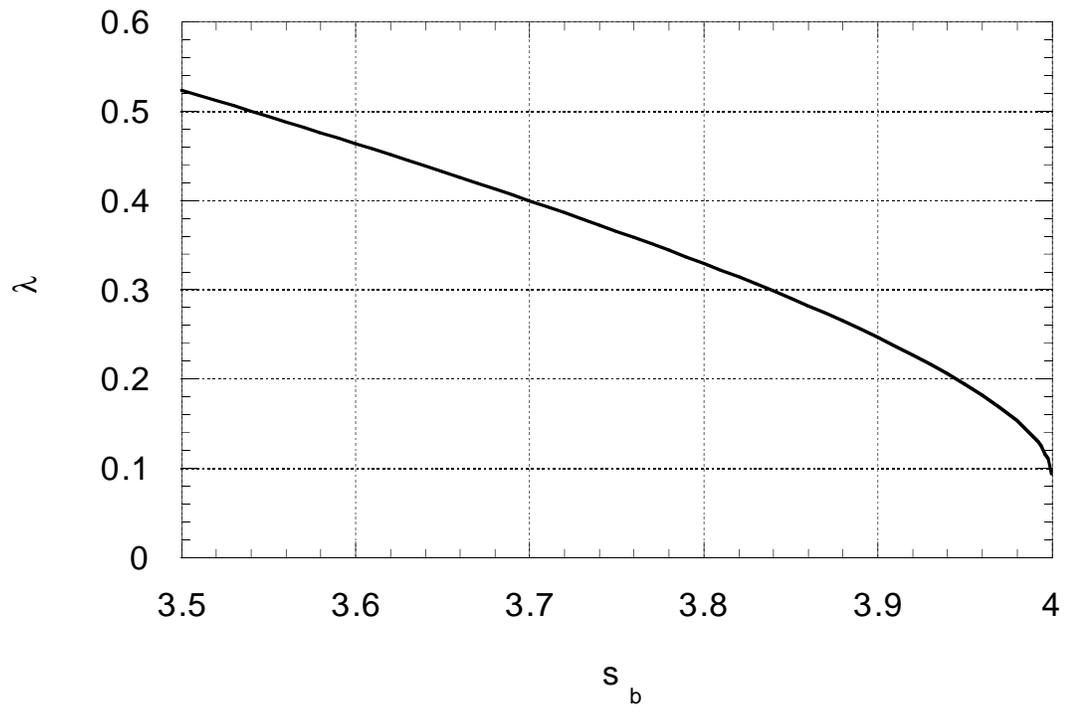

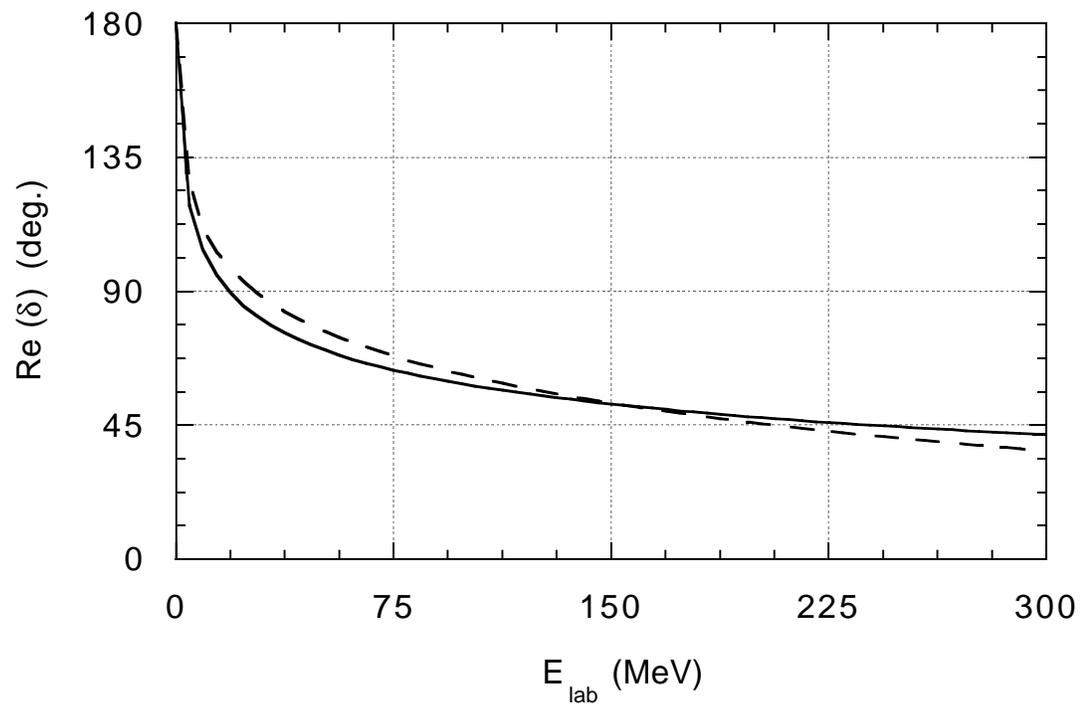

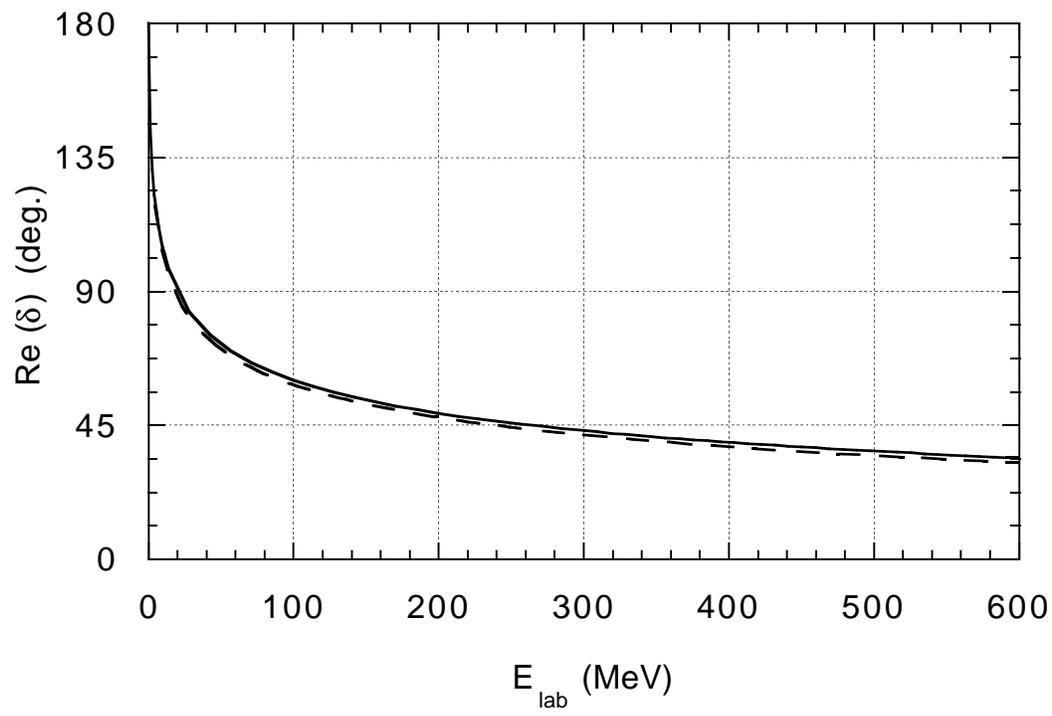

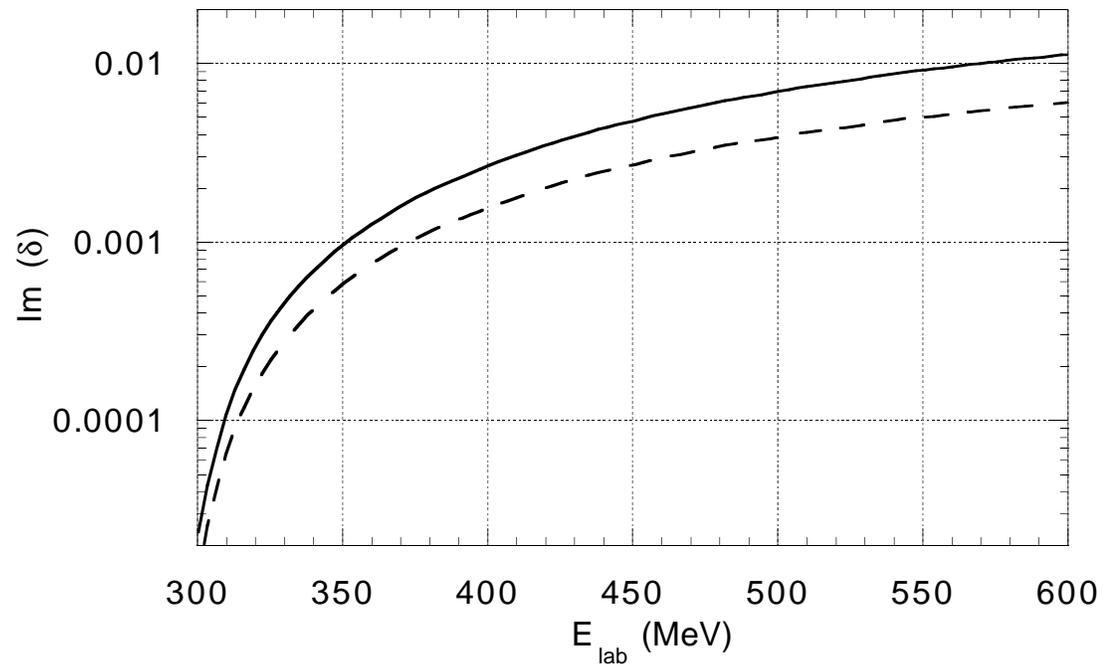

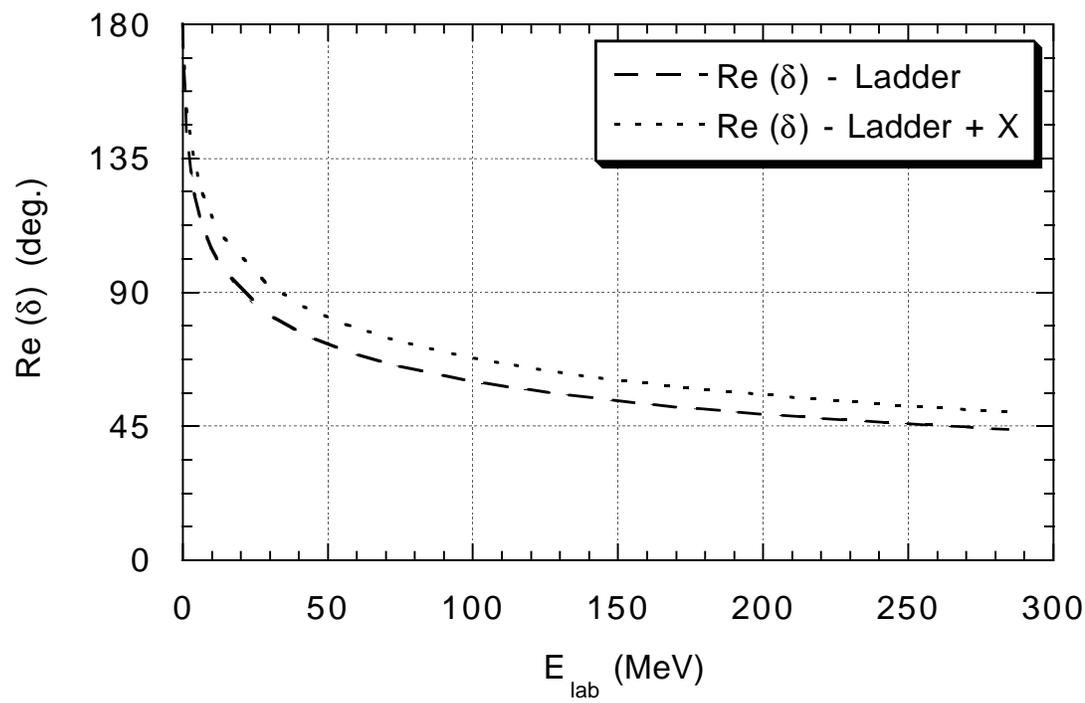

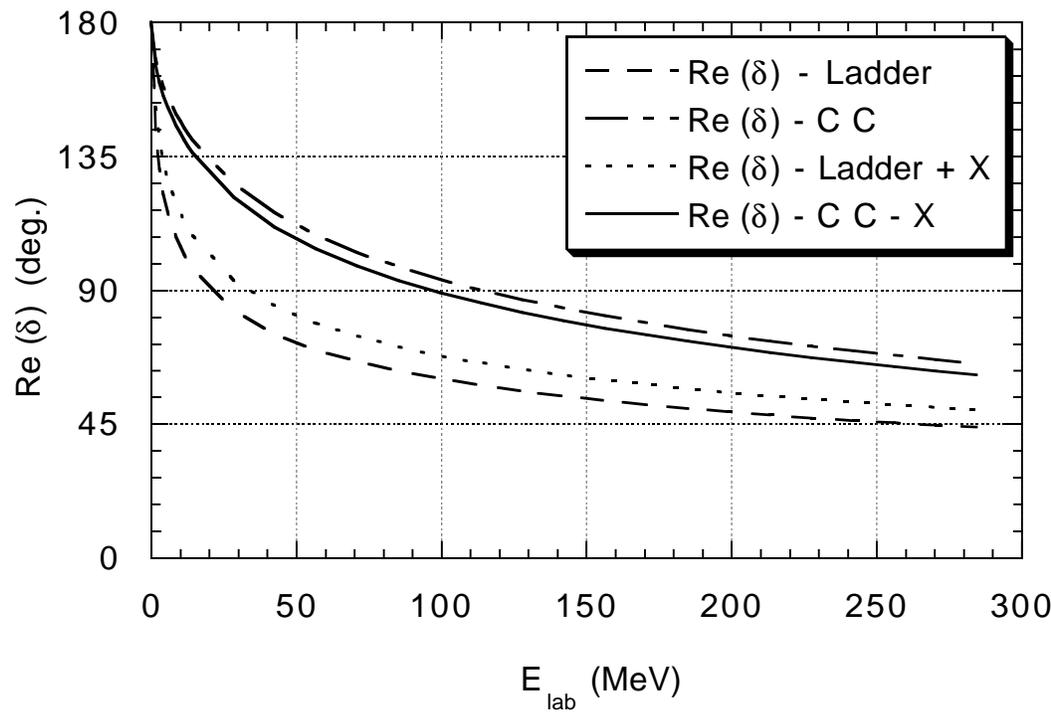

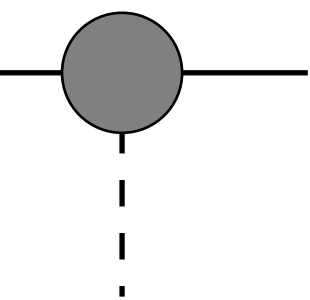 = 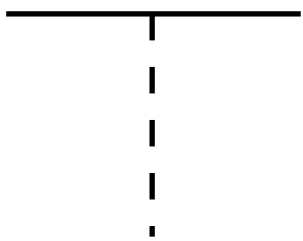 + 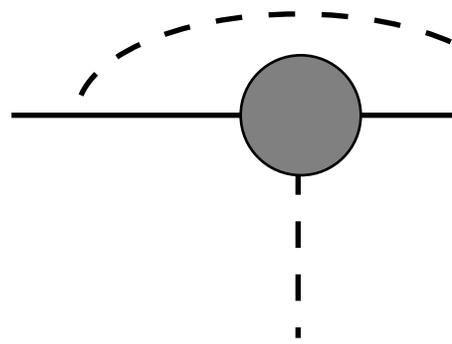

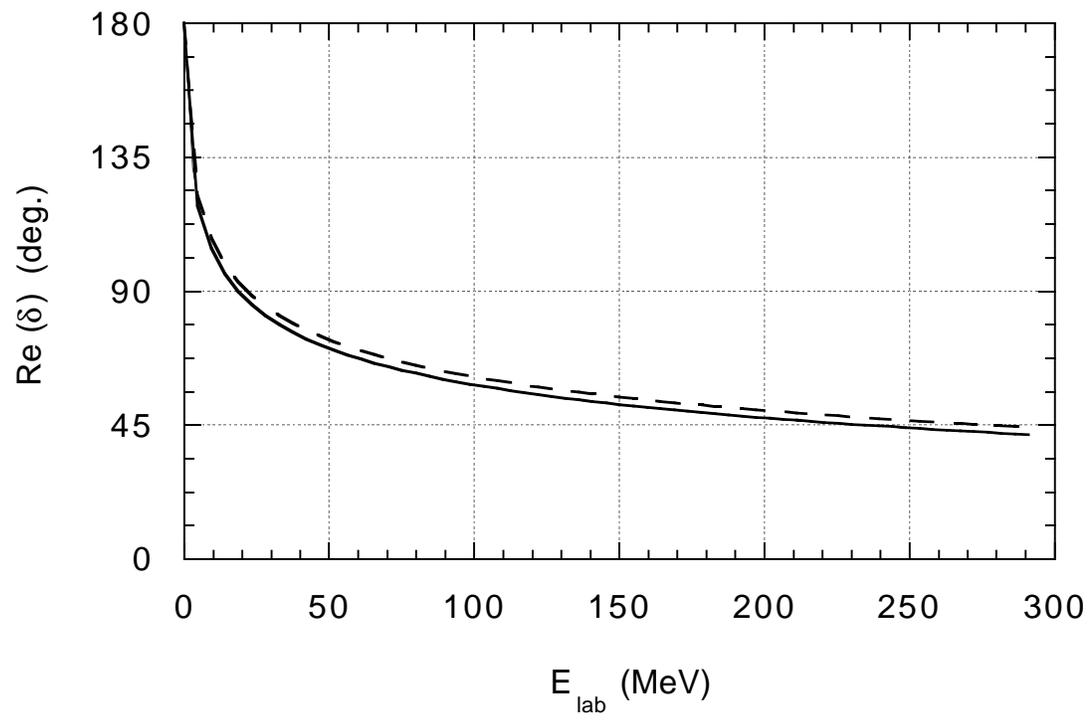

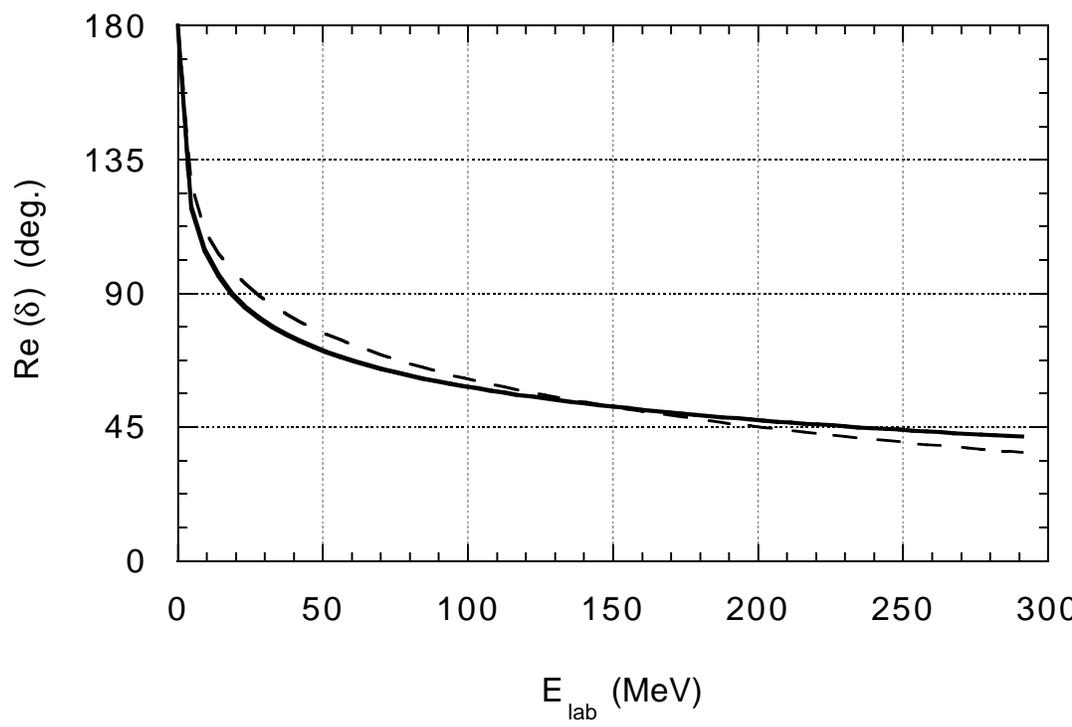

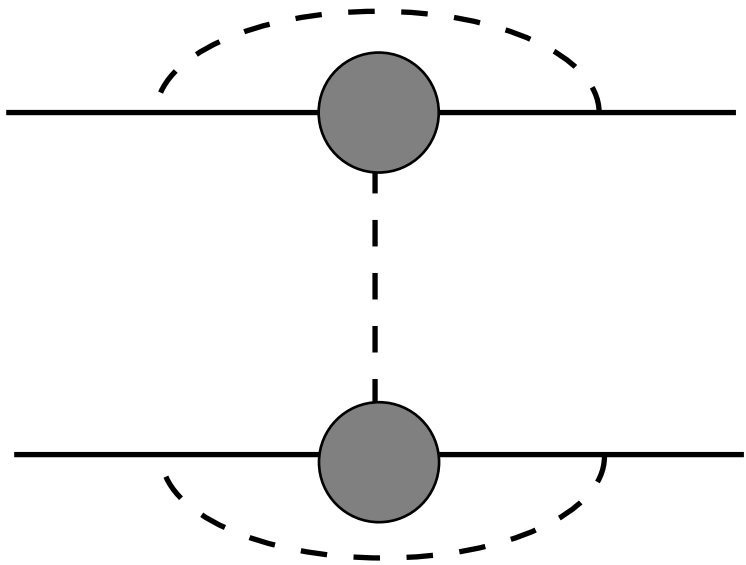

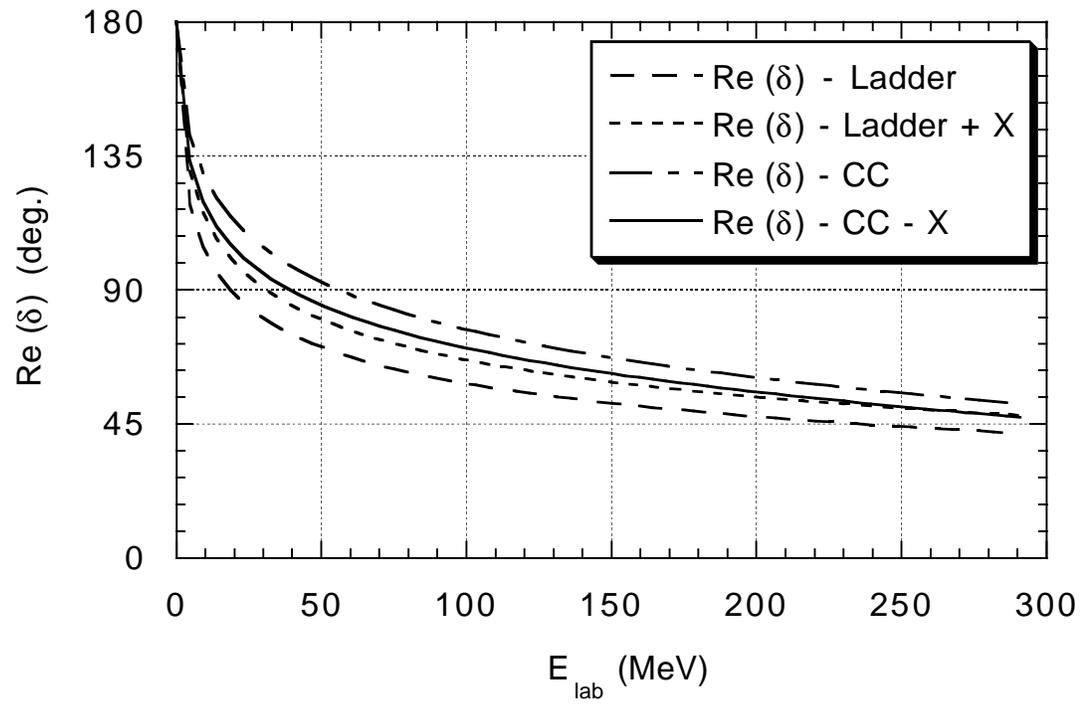

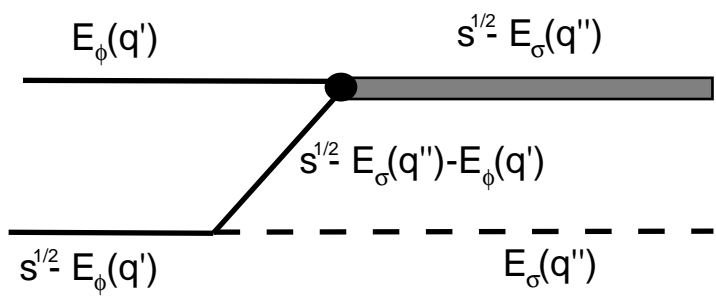

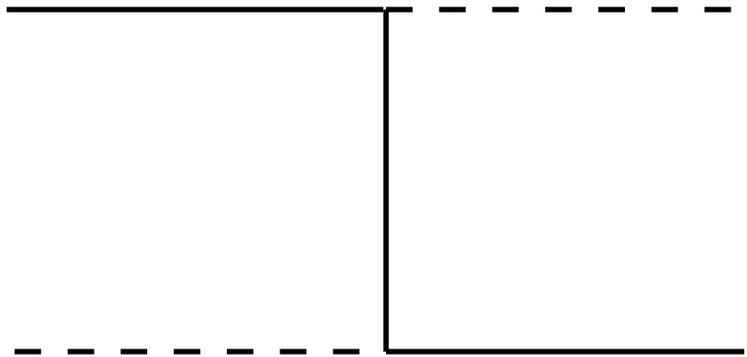